\documentstyle[epsfig,12pt,preprint,tighten,aps]{revtex}
\begin{document}

\draft

\title{\rightline{{\tt June 1999}}
\vskip 1cm
%\rightline{{\tt UM-P-99/xx}}
%\rightline{{\tt RCHEP-99/xx}}
\ \\
Detailed study of BBN implications of neutrino oscillation
generated neutrino asymmetries in some four
neutrino models}
\author{R. Foot}
\address{School of Physics\\
Research Centre for High Energy Physics\\
The University of Melbourne\\
Parkville 3052 Australia\\
(foot@physics.unimelb.edu.au)}

\maketitle

\begin{abstract}
We re-examine the evolution of neutrino asymmetries in
several four neutrino models. The first case
involves the direct creation of $L_{\nu_e}$ by
$\nu_e \leftrightarrow \nu_s$ oscillations.
In the second case, we consider the mass hierarchy
$m_{\nu_\tau} \gg m_{\nu_\mu}, m_{\nu_e}, m_{\nu_s}$
where $\nu_\tau \leftrightarrow \nu_s$ oscillations
generate a large $L_{\nu_\tau}$ and some of this asymmetry
is converted into $L_{\nu_e}$ by $\nu_{\tau} 
\leftrightarrow \nu_{e}$ oscillations.
We estimate the implications for BBN for a
range of cosmologically interesting $\delta m^2$ values.
The present paper improves on previous published
work by taking into account the finite repopulation
rate and the time dependence of the distortions
to the neutrino momentum distributions.
The treatment of chemical decoupling is also improved.

\end{abstract}

\newpage
\section{Introduction}

If light sterile neutrinos exist, then this will lead
to important implications for early Universe cosmology.
This is because ordinary-sterile neutrino oscillations
generate large neutrino asymmetries for a large range
of parameters\cite{ftv,fv1,fv2,f,bvw}. This is a generic feature 
of ordinary-sterile neutrino oscillations.

The implications of this phenomena are quite model
dependent.  Various implications of this phenomena have been
discussed in a number of previous papers for
a number of interesting models motivated
by the existing neutrino anomalies\cite{fv1,fv2,f,bfv,fv3,fv4}.
For example in Refs.\cite{fv1,f} it has been shown that
the maximal $\nu_\mu \to \nu_s$ oscillation solution  
to the atmospheric neutrino anomaly is consistent with
a stringent BBN bound of $N^{BBN}_{eff} \stackrel{<}{\sim} 
3.6$ (and may also be consistent with $N^{BBN}_{eff} < 3$
depending on the model\cite{fv2,fv4}).
This consistency requires $m_{\nu_\tau} \stackrel{>}{\sim}
{\rm few}\ eV$ (for $|\delta m^2_{atmos}| \simeq 3\times 10^{-3}
eV^2$) thus placing the $\nu_\tau$ in the interesting
hot dark matter range.
Of course this is also of great interest to short-baseline
experimentalists!

Of particular concern to this paper, is the `low temperature'
evolution of neutrino asymmetries which also has important
implications for BBN.
In Ref.\cite{fv2}, we discussed the four neutrino model with
$m_{\nu_\tau} \gg m_{\nu_\mu}, m_{\nu_e}, m_{\nu_s}$. 
In this case a large $L_{\nu_\tau}$ asymmetry
is generated by $\nu_\tau \leftrightarrow \nu_s$ 
oscillations, some of which is transferred to 
$L_{\nu_e}$ by $\nu_\tau \leftrightarrow \nu_e$ oscillations.
This has important implications for BBN 
since it allows $N^{BBN}_{eff} < 3$, with
$N^{BBN}_{eff} \approx 2.5$ for a large range
of parameters if $L_{\nu_e} > 0$. 
Qualitatively similar results occur 
for other sterile neutrino models as has been shown in
a number of recent papers\cite{bfv,fv4}. 
One point of all this is that $N^{BBN}_{eff} < 3$
is a serious possibility if light effectively sterile neutrinos
exist.

In the paper Ref.\cite{fv2} we considered the case of large
$|\delta m^2|$ where
the $L_{\nu_e}$ was created above 1.5-2.0 MeV, so that
its implications for BBN could be approximately
discussed by using thermal neutrino distributions (i.e.
the neutrino asymmetry was distributed 
with chemical potentials) which were approximately
constant during the
era when the neutron/proton ratio 
was changing  (i.e. for $T \stackrel{<}{\sim} 2\ MeV$).
We also briefly estimated the effects for the
direct production of $L_{\nu_e}$ by 
$\nu_e \to \nu_s$ oscillations.
(this case is only possible if the other neutrinos 
are lighter or do not oscillate into the sterile neutrino).

Recently, Ref.\cite{fv2} has been criticised in Ref.\cite{sfa}
where it is claimed that the time dependence
of the neutrino asymmetry and finite repopulation
rate (which was assumed to be instantaneous 
in Ref.\cite{fv2} for temperatures above about 1.5 MeV)
is of critical importance.  
Ref.\cite{sfa} also similarly criticises Ref.\cite{bfv} 
(which studied a quite different 4 neutrino model with
approximately degenerate $\nu_\mu, \nu_\tau$ states)
but this is clearly unjustified because
Ref.\cite{bfv} takes into account the finite
repopulation rate using a Pauli-Boltzman approach
(as well as the time dependence of the distortion
in the neutrino distributions). In fact,
Ref.\cite{sfa} appears to follow the repopulation
procedure advocated in Ref.\cite{bfv} and
re-examines the cases in Ref.\cite{fv2} using this
repopulation procedure. 
In view of this, we have also decided to revisit the models
considered in Ref.\cite{fv2} in this paper
because we believe that the results of Ref.\cite{sfa}
to be incorrect. 
We will compute the evolution of the number distributions
taking into account the finite repopulation rate and time
dependence of the asymmetry.
As already emphasised above,
such an approach was already used in the 
papers\cite{bfv,fv4} discussing different models,
so it is straightforward to apply it here.
We will also improve on Ref.\cite{fv2} by discussing
more completely the effects of the two similar
oscillation modes, $\nu_\tau \leftrightarrow \nu_\mu,
\ \nu_\tau \leftrightarrow \nu_e$.
We also give a more accurate treatment of the kinetic
decoupling region which suggests a slightly lower
kinetic decoupling temperature.

\section{Big bang nucleosynthesis }

The primordial deuterium to hydrogen ($D/H$) 
ratio can be used to give a sensitive
determination of the baryon to photon ratio $\eta$ 
which, given the estimated primordial
$^4$He mass fraction, can be used to infer the 
effective number of light neutrino flavours
$N_{eff}^{BBN}$ during the BBN epoch. This value 
can then be compared with the
predictions for $N_{eff}^{BBN}$ from various models 
of particle physics to find out
which ones are compatible with standard BBN. 
For example, the minimal standard model
predicts $N_{eff}^{BBN} = 3$. At the present time, most 
estimates favour $N_{eff}^{BBN} < 3.6$ and some
estimates favour $N_{eff}^{BBN} < 3.0$\cite{bbn1}.  
Of course, even if a model of particle physics is shown to be
incompatible with BBN, this does not necessarily mean that 
the model is incorrect, since for example, it 
is also possible that one of the standard assumptions of 
BBN may not be correct\cite{ka}.

For gauge models with effectively 
sterile neutrinos, one in general expects
$N_{eff}^{BBN} \neq 3$.
In fact, $N_{eff}^{BBN}$ may be less than three or greater than
three.  The prediction for $N_{eff}^{BBN}$ depends on the 
oscillation parameters in a given model and also
the sign of the asymmetry (which for various reasons
cannot be predicted at the moment).
One possible consequence of ordinary-sterile
neutrino oscillations is the
excitation of sterile neutrino states, which 
typically leads to an increase in the expansion rate of
the universe and thereby also increases $N_{eff}^{BBN}$.
Another possible consequence of ordinary-sterile
neutrino oscillations is the
dynamical generation of an electron-neutrino asymmetry. 
This also has important implications for BBN,
as it directly
affects the reaction rates which determine the neutron to proton ($n/p$) 
ratio just before nucleosynthesis.
If the electron neutrino asymmetry is positive then it will
decrease $N_{eff}^{BBN}$, while if it
is negative then it will increase $N_{eff}^{BBN}$.
 
The neutron to nucleon ratio, $X_n (t)$, is related to 
the primordial Helium mass fraction,
$Y_P$, by\footnote{For a review of helium synthesis, see for example
Ref.\cite{weinberg}.}
\begin{equation}
Y_P = 2X_n
\end{equation}
just before nucleosynthesis.
The evolution of $X_n (t)$ is governed by the equation,
\begin{equation}
\frac{dX_n}{dt}=-\lambda(n\rightarrow p)X_n +
\lambda(p\rightarrow n)(1-X_n),
\end{equation}
where the reaction rates are approximately
\begin{eqnarray}
\label{eq:rates}
\lambda(n\rightarrow p) = \lambda(n + \nu_e \rightarrow p + e^-) 
+ \lambda(n + e^+ \rightarrow p + \overline{\nu}_e)
+ \lambda (n \rightarrow p + e^- + \overline{\nu}_e), \nonumber \\
\lambda(p\rightarrow n) = \lambda(p + e^- \rightarrow n + \nu_e) 
+ \lambda(p + \overline{\nu}_e \rightarrow n + e^+) +
\lambda(p + e^- + \overline{\nu}_e \rightarrow n).
\end{eqnarray}
These reaction rates 
depend on the momentum distributions of the species involved.
The 2-body processes in Eq.(\ref{eq:rates}) for 
determining $n\leftrightarrow p$ are only important
for temperatures above about $0.4$ MeV. Below this temperature these 
weak interaction rates
freeze out and neutron decay becomes the dominant 
factor affecting the $n/p$ ratio.  For example,
an excess of $\nu_e$ over $\overline{\nu}_e$, due
to the creation of a positive $L_{\nu_e}$ would 
change the rates for the processes in Eq.(\ref{eq:rates}). 
The effect of this would be to reduce the n/p ratio, and hence
reduce $Y_P$.
Neutron decay is not significantly altered by 
lepton asymmetries unless they are very large.   
It is quite well known that
a small change in $Y_P$ due to the modification of 
$\nu_e$ and $\overline \nu_e$ distributions
does not impact significantly on the 
other light element abundances (see for example Ref.\cite{olive}).
A small modification to the expansion rate, 
using the convenient unit $N_{eff}^{BBN}$, 
primarily affects only $Y_P$, with\cite{walker}
\begin{equation}
\delta Y_P \simeq 0.012\times \delta N_{eff}^{BBN}.
\label{walk}
\end{equation}
Since Appendix A of Ref.\cite{fv4}, describes 
in detail how we compute the effect
on $Y_P$ due to the modified $\nu_e$ and $\overline \nu_e$ distributions,
we will not waste any more paper by discussing it further here.

\section{Case 1: Implications for BBN of $\nu_e \leftrightarrow \nu_s$
oscillation generated $L_{\nu_e}$}

In this section we will study the direct production of
a large $L_{\nu_e}$ from $\nu_e \leftrightarrow \nu_s$
oscillations. We will ignore oscillations involving
 $\nu_\mu$ or $\nu_\tau$. 
This is only an approximately valid
thing to do provided that either their masses are very small 
(so that the largest $|\delta m^2|$ belongs to the
$\nu_e \leftrightarrow \nu_s$ oscillations and
the other oscillations have $|\delta m^2|$ much
less than $1 \ eV^2$) {\it or } that
they do not mix with the $\nu_e, \nu_s$ (i.e. the
$\nu_e, \nu_s$ decouple from the $\nu_\mu, \nu_\tau$ in
the neutrino mass matrix).

Let us begin with some necessary preliminaries.
Our notation/convention for ordinary-sterile 
neutrino two state mixing is as follows. The weak
and sterile eigenstates $\nu_{\alpha}$ ($\alpha = e, \mu, \tau$) 
and $\nu_s$ are linear combinations
of two mass eigenstates $\nu_a$ and $\nu_b$,
\begin{equation}
\nu_{\alpha} = \cos\theta_{\alpha s} \nu_a + 
\sin\theta_{\alpha s} \nu_b,\
\nu_{s} = - \sin\theta_{\alpha s} \nu_a + 
\cos\theta_{\alpha s} \nu_b,
\label{zwig}
\end{equation}
where $\theta_{\alpha s}$ is the vacuum mixing angle.
We define $\theta_{\alpha s}$ so that 
$\cos2 \theta_{\alpha s} > 0$ and
we adopt the convention that $\delta m^2_{\alpha s} 
\equiv m^2_b - m^2_a$.

Recall that the $\alpha$-type neutrino asymmetry is defined by 
\begin{equation}
L_{\nu_\alpha} \equiv 
{n_{\nu_\alpha} - n_{\overline \nu_\alpha} \over n_\gamma}.
\label{def}
\end{equation}
In the above equation, $n_{\gamma}$ is the
number density of photons, $n_{\gamma} = 2\zeta (3)T^3/\pi^2$.

Note that when we refer to ``neutrinos'', sometimes we
will mean neutrinos and/or antineutrinos.
We hope the correct meaning will be clear from context.
Also, if neutrinos are Majorana particles, then technically
they are their own antiparticle. 
Thus, when we refer to ``antineutrinos'' we obviously
mean the right-handed helicity state in this case.

In Ref.\cite{ftv} it was shown that ordinary-sterile
neutrino oscillations generate
large neutrino asymmetries for a wide range of parameters.
This work built upon earlier work on ordinary-sterile
neutrino oscillations in the early Universe\cite{early}.
As already discussed in detail in previous publications
\cite{fv2,fv1,ftv}
the evolution of lepton number can be separated into
three distinct phases. 
At high temperatures the oscillations are damped
and evolve so that $L^{(\alpha)} \to 0$ (where
$L^{(\alpha)} \equiv L_{\nu_\alpha} + L_{\nu_e} + L_{\nu_\mu}
+ L_{\nu_\tau} + \eta$, and $\eta$ is related to the baryon
asymmetry).
In this region the resonance momentum for neutrino oscillations
is approximately the same as anti-neutrino oscillations.
If $\delta m^2_{\alpha s} < 0$ then
at a certain temperature, $T_c$, which is given
roughly by\cite{ftv},
\begin{equation}
T_c \sim 
16\left({-\delta m^2_{\alpha s} \cos2\theta_{\alpha s} \over 
\text{eV}^2}\right)^{1 \over 6}\ \text{MeV},
\end{equation}
exponential growth of neutrino asymmetry occurs
(which typically generates a neutrino asymmetry
of order $10^{-5}$ at $T\simeq T_c$, see figure 1 of Ref.\cite{f}
for some typical examples).
Taking for definiteness that the $L_{\nu_\alpha}$ is
positive, the anti-neutrino oscillation resonance
moves to very low values of $p/T \sim 0.5$
while the neutrino oscillation resonance moves
to high values $p/T \stackrel{>}{\sim} 10$
(see Ref.\cite{fv2} for a figure illustrating this).
The subsequent evolution of neutrino asymmetries,
which is dominated by adiabatic MSW transitions of
the antineutrinos, follows an orderly $1/T^4$ behaviour
until the resonance has passed through the entire distribution. 
The final asymmetry generated is typically in the 
range $0.23 \stackrel{<}{\sim} L_{\nu_\alpha} \stackrel{<}{\sim}
0.37$\cite{fv2}.
Because the oscillations are dominated by adiabatic
MSW behaviour it is possible to use a relatively
simple and accurate formalism to describe the 
evolution of the system at the `low temperatures',
$T \stackrel{<}{\sim} T_c/2$.
In fact, we only need to know the values of
the oscillation resonance momentum at $T \simeq T_c/2$.
Previous numerical work has already
shown\cite{fv2} that by $T \simeq T_c/2$, neutrino asymmetry
is generated such that $0.2 
\stackrel{<}{\sim} p/T \stackrel{<}{\sim} 0.8$
(the precise value depends on $\sin^2\theta_{\alpha s}, 
\delta m^2_{\alpha s}$).
Furthermore the subsequent evolution is approximately
insensitive to the initial value of $p/T$ in this range
(provided, of course, that negligible number
of sterile neutrinos were produced at high temperature).

In this section we will deal with the case of
$\nu_e \leftrightarrow \nu_s$ oscillations
directly producing the $L_{\nu_e}$ asymmetry.
For the implications for BBN we are primarily
interested in the ``low temperature'' evolution of the
number distributions and lepton numbers in this case.
Our analysis can be broken up into the following steps:

\begin{enumerate} 

\item We assume complete adiabatic MSW conversion
of neutrinos at the MSW resonance.

\item From this we can compute the evolution of 
lepton number asymmetries which 
not only dictates the momentum of the MSW resonances,
but also the chemical potentials.

\item Using these chemical potentials we can evaluate
the equilibrium distributions from which we
can estimate the actual distributions by a
Pauli-Boltzman repopulation equation. 
\end{enumerate}
We now discuss each of these steps in detail.

Consider, for the moment, two-flavour small angle 
(i.e. $\cos 2\theta_{es} \simeq 1$) ordinary-sterile
$\nu_e \leftrightarrow \nu_s$ neutrino oscillations. 
As discussed in detail in earlier papers
the $\nu_e \leftrightarrow \nu_s$ neutrino oscillations
only generate $L_{\nu_e}$ provided that 
$\delta m^2_{es} < 0$ and this will be
assumed in the forthcoming discussion\footnote{
Note that $|\delta m^2_{es}| \le m^2_{\nu_e}$  
(where $m_{\nu_e}$ is the mass of the state which
is predominately $\nu_e$). 
Recall that there is an experimental upper bound 
on $m_{\nu_e}$ which is a few eV if $\nu_e$ is 
a Dirac neutrino and about an eV if $\nu_e$ 
is a Majorana neutrino\cite{pdg}.}.
We know from numerical integration of the
exact quantum kinetic equations\cite{fv2} that the
adiabatic approximation is
valid provided that $\sin^2 2\theta_{e s} \stackrel{>}{\sim} 
{\rm few} \times 10^{-10}$.  Now, coherent small angle
adiabatic MSW transitions completely convert 
$\nu_e \leftrightarrow    
\nu_s$ at the resonance momentum of these states.  
The resonance momentum is given approximately by 
(see e.g. Ref.\cite{fv2}),
\begin{equation}
{p_{res} \over T} = {|\delta m^2_{e s}| \over 
a_0 T^4 L^{(e s)}},
\label{7}
\end{equation}
where $a_0 \equiv 4\sqrt{2}\zeta(3)G_F/\pi^2$ and 
\begin{equation}
L^{(e s)} \equiv  2L_{\nu_e} + L_{\nu_\tau}.
\label{9}
\end{equation}
In the above equation we have neglected the $L_{\nu_\mu}$
asymmetry (as well as the baryon/electron asymmetry). This is because
these asymmetries are unimportant in the low temperature
region unless they happen to be large (i.e. 
greater than about $10^{-5}$).
For adiabatic two-flavour neutrino oscillations in the 
early universe it is quite easy to see that the 
rate of change of lepton number is 
governed by the simple equation\cite{fv2},
\begin{equation}
{dL_{\nu_e} \over dT} = 
-X(p_{\text{res}})\left| {d(p_{\text{res}}/T) \over dT}\right|,
\label{he}
\end{equation}
where $p_{\text{res}}$ is the MSW resonance momentum,
Eq.(\ref{7}) and
\begin{equation}
X(p) = 
{T \over n_\gamma}\left[ N_{\bar \nu_e}(p) - N_{\bar \nu_s}(p)
\right].
\label{he6}
\end{equation}
Also, $N_{\bar \nu_e}(p)$ and $N_{\bar \nu_s}(p)$
are the momentum distributions of the $\overline \nu_e$ 
and $\overline \nu_s$ states.
In the above equations, the case $L_{\nu_e} > 0$ 
has been considered
(so that the resonance occurs for antineutrinos).
Equation (\ref{he}) relates the rate of change 
of lepton number to the speed of the
resonance momentum through the neutrino distribution. 
Reference \cite{bvw} provides a
detailed discussion of how this equation can be derived from 
the quantum kinetic equations for the case of
adiabatic evolution with a narrow resonance 
width.  As discussed in Ref.\cite{fv2}, 
Equation (\ref{he}) can be simplified using 
\begin{equation}
{d(p_{\text{res}}/T) \over dT} = {\partial (p_{\text{res}}/T)
\over \partial T} + {\partial (p_{\text{res}}/T) \over 
\partial L_{\nu_e}} {dL_{\nu_e} \over dT},
\label{presderiv}
\end{equation}
 from which it follows that
\begin{equation}
{dL_{\nu_e} \over dT} = 
{fX{\partial (p_{\text{res}}/T) \over \partial T}
\over 1 - fX{\partial (p_{\text{res}}/T) \over \partial L_{\nu_e}}}  
\ = \ {-4fXp_{\text{res}}\over
T^2 + {2 f TX p_{\text{res}} \over L^{(es)}}},
\label{presevol}
\end{equation}
where $f = 1$ for 
$d(p_{\text{res}}/T)/dt > 0$ 
(that is for $d(p_{\text{res}}/T)/dT < 0$) 
and $f = -1$ for
$d(p_{\text{res}}/T)/dt < 0$ and
we have dropped the momentum dependence of $X$ in 
the above equation for notational clarity\footnote{
At $T = T_c/2, f=1 $ and it does not change sign
during subsequent evolution.}.  
Eq.(\ref{presevol}) allows us to compute the evolution of
$L_{\nu_e}$. As discussed earlier, it is valid from
$T \simeq T_c/2$ (with $p_{res}/T \sim 0.5$ at this
point). For the more
complicated multi-flavour case considered in section IV, 
coupled equations based on Eq.(\ref{presevol}) will be used. 
We now must describe how we compute the evolution of
$N_{\nu_\alpha}$.

The MSW transitions effect the adiabatic conversion
\begin{equation}
|\overline \nu_e \rangle \leftrightarrow 
|\overline \nu_s\rangle. 
\end{equation}
This means that as $P_1$ sweeps through the $\overline \nu_e$
momentum distribution, 
\begin{eqnarray}
N_{\overline \nu_s}(P_1) &\to& N_{\overline \nu_e}(P_1), \nonumber \\
N_{\overline \nu_e}(P_1) &\to& N_{\overline \nu_s}(P_1).
\label{p1}
\end{eqnarray}
In our numerical work the continuous momentum 
distribution for each flavour is replaced by a 
finite number of `cells' on a logarithmically
spaced mesh. As the momentum $P_1$ passes
a cell, the number density in the cell
is modified according to Eq.(\ref{p1}).
Of course weak interactions will repopulate
some of these cells as they thermalise the
neutrino momentum distributions.
The repopulation can also generate small
chemical potentials for the other flavours
(as will be discussed latter).
We take repopulation into account with the rate equation
for each flavour $\alpha = e, \mu, \tau$,
\begin{eqnarray}
{\partial \over \partial t} {N_{\nu_\alpha}(p) \over N_{0}(p,T)}
&\simeq & \Gamma_{\alpha}(p)
\left[{N^{\text{eq}}(p,T,\mu_{\nu_{\alpha}}) \over N_{0}(p,T)}   
- {N_{\nu_\alpha}(p) \over N_{0}(p,T)} \right],
\nonumber \\
{\partial \over \partial t} {N_{\overline \nu_\alpha}(p) 
\over N_{0}(p,T)}
&\simeq & \Gamma_{\alpha}(p)
\left[{N^{\text{eq}}(p,T,\overline \mu_{\nu_{\alpha}}) 
\over N_{0}(p,T)}   
- {N_{\overline \nu_\alpha}(p) \over N_{0}(p,T)} \right],
\label{dfd}
\end{eqnarray}
where $\Gamma_{\alpha}(p)$ is the total collision
rate and is approximately given by 
\begin{equation}
\Gamma_{\alpha}(p) = y_{\alpha}G_F^2T^5\left({p \over 3.15T}\right),
\end{equation}
with $y_e \simeq 4.0, y_{\mu,\tau} \simeq 2.9$ and $G_F$
is the Fermi constant.
Also, in Eq.(\ref{dfd}), $N_{0}(p,T), 
\ N^{eq}(p,T,\mu_{\nu_\alpha})$ are the
equilibrium distributions with zero chemical
potential and chemical potential $\mu_{\nu_\alpha}$
respectively\footnote{
Our convention for the sign of the chemical potential
is $N^{eq}(p,T,\mu_{\nu_{\alpha}}) = {1 \over 2\pi^2}
{p^2 \over 1 + e^{(p+\mu_{\nu_{\alpha}})/T}}$.}.  
Previous papers\cite{bfv,fv4} used a simple approximation 
whereby the transition out of chemical equilibrium
occurred at the decoupling temperature $T_{dec}^{\alpha}$.
Obviously this is not a sharp transition.
Also, there will be small chemical
potentials created by the other flavours
as they create $\overline \nu_e \nu_e$ pairs
to compensate for the loss of $\overline \nu_e$ states.
In the appendix we discuss a more 
accurate (but more complicated) formalism 
to compute the values of the chemical potentials
for all of the flavours as a function of time (or 
equivalently temperature). 
The conclusion is that the simple treatment 
of chemical decoupling, discussed in
previous papers is roughly valid, but
noticable differences (though typically 
not greater than about $\delta N^{BBN}_{eff} \sim 0.2$) 
for BBN can occur.
[Although there will not be much difference for the
$\nu_e \leftrightarrow \nu_s$ case since the lepton
number is generated so late, since experimentally
$|\delta m^2_{es}| \stackrel{<}{\sim} 10\ eV^2$ (see
footnote 2)].  We will use the more complicated 
formalism discussed in the appendix to evaluate the
chemical potentials for all of our numerical work in
this paper.

Using the above procedure the lepton asymmetry $L_{\nu_e}$,
and the neutrino distributions,
$N_{\nu_\alpha}(p,t),\ N_{\overline \nu_\alpha}(p,t)$ can 
be obtained. We can feed the $N_{\nu_e}(p,t),\ 
N_{\overline \nu_e}(p,t)$
distributions into a nucleosynthesis code 
(which we integrate concurrently) in order
to compute the implications for BBN. 
It is useful to separate the total contribution 
to $\delta Y_P$ into two contributions,
\begin{equation}
\delta Y_P = \delta_1 Y_P + \delta_2 Y_P,
\end{equation}
where $\delta_1 Y_P$ is the change due
to the effect of the modified electron neutrino 
momentum distributions on the nuclear reaction rates, and
$\delta_2 Y_P$ is
due to the change in the energy density (or equivalently
the change in the expansion rate of the universe).
While BBN is only sensitive (to a good approximation)
to the total contribution, $\delta Y_P$, the
separate parts will have quite different implications
for the forthcoming precision measurements of the
anisotropy of the cosmic microwave background.
In particular it may be possible to
estimate the expansion rate of the Universe at
the time of photon decoupling\cite{raf}.
 
The contribution $\delta_1 Y_P$ can be determined by  
numerically integrating the rate equations 
for the processes given in Eq.(\ref{eq:rates}) 
using the modified electron neutrino momentum 
distributions $N_{\nu_e}$ and $N_{\overline \nu_e}$
as discussed in Appendix A of Ref.\cite{fv4}. 
The contribution $\delta_2 Y_P$ can be computed
from the momentum distributions of the ordinary and
sterile neutrinos through
\begin{equation}
\delta_2 Y_P \simeq 0.012 \left(
{1 \over 2\rho_0} 
\int_0^{\infty} \left[N_{\overline \nu_s}(p) 
+ \sum_{\alpha =1}^3 N_{\nu_\alpha}(p) +
N_{\overline \nu_\alpha}(p) 
\right]pdp - 3\right),
\end{equation}
where 
\begin{equation}
\rho_0 \equiv \int^{\infty}_0 N_{0}(p,T) pdp 
= {7 \pi^2 \over 240} T^4,
\label{Weylrho}
\end{equation}
is the energy density of a Weyl fermion at equilibrium with 
zero chemical potential.  [Recall that Eq.(\ref{walk}) 
can be used to express $\delta Y_P$, $\delta_1 Y_P$ and 
$\delta_2 Y_P$ in terms of effective neutrino number,
$\delta N_{eff}^{BBN}$, $\delta_1 N_{eff}^{BBN}$ and $\delta_2    
N_{eff}^{BBN}$, respectively.]
To calculate $\delta_2 Y_P$, we numerically determine the 
momentum distributions at $T = 0.5\ \text{MeV}$. 
Because of the approximate kinetic decoupling of neutrinos
for temperatures below about $3$ MeV,
large contributions\footnote{By `large
contributions' we mean $\delta_2 N_{eff}^{BBN}
\stackrel{>}{\sim} 0.10$.} to $\delta_2 Y_P$, should they exist,
must have been generated earlier. A temperature of 
$0.5\ \text{MeV}$ is therefore a safe
place to evaluate the final $\delta_2 Y_P$.

Recall that there is an ambiguity concerning the sign 
of the $L_{\nu_e}$ lepton
asymmetry. We have considered the $L_{\nu_{e}} > 0$ case above
for definiteness, 
but $L_{\nu_e} < 0$ is equally likely a priori. 
Previous work Ref.\cite{fv1} has shown that the sign
is fixed in the region where the 'static approximation'
is valid. This approximation assumes that the 
the asymmetry evolution is 
dominated by collisions and is sufficiently smooth 
\footnote{
In fact, it has been shown in Ref.\cite{bvw} that 
this approximation is equivalent to the adiabatic limit
of the quantum kinetic equations in the region
where collisions dominate the evolution of the neutrino asymmetry.}. 
Importantly it is generally valid in
the region $T \simeq T_c$ where the neutrino asymmetry
is initially generated\cite{fv1} provided that 
$\sin^2 2\theta \stackrel{<}{\sim} 
{\cal O}\left( 10^{-6} \right)$
\cite{fv1} for $\delta m^2 \sim -10 \ eV^2$. 
Thus in this region of parameters the sign is fixed.
For $\sin^2 2\theta \stackrel{>}{\sim} {\cal O} \left(10^{-6}\right)$
numerical integration of the quantum kinetic equations
(including the momentum distribution of the neutrinos)
reveals\cite{f} that the sign oscillates
\footnote{
Note that the recent study in Ref.\cite{rec}, which
neglects the neutrino momentum distribution 
arrives at quite different results.
They find that the region where the sign doesn't oscillate
is much smaller, roughly, $\sin^2 2\theta \stackrel{<}{\sim} 
{\cal O} \left(10^{-8}\right)$ for $\delta m^2 \sim -10 \ eV^2$.
Qualitatively, this is very easy to understand.
This reason is that in the average momentum toy-model,
all of the neutrinos enter the MSW resonance 
at the same time which significantly enhances
the rate at which neutrino asymmetry is created at
$T = T_c$.
The rapid creation of neutrino asymmetry reduces
the region where the oscillations are adiabatic\cite{fv1}.
Also, the paper in Ref.\cite{shi} similarly
assumes that all of the neutrinos have the same
momentum, but (perhaps not suprisingly) obtain
quite different results to Ref.\cite{rec}.}
for a short period at $T \simeq T_c$.
In the parameter region where the sign
oscillates (and possibly in some of the
parameter region where it doesn't oscillate)
the sign may not be fixed when fluctuations
are considered\cite{pd}.
It may be possible for different regions of space
to have different signs of the lepton number, as
first suggested in Ref.\cite{ftv}.
Whether this happens or not is an open question
at the moment and will depend on the size of the 
fluctuations present.

In any case, even if the sign is fixed, it 
cannot be predicted as it
depends on the initial values of the neutrino asymmetries
(as well as the baryon asymmetry).
For the negative $L_{\nu_{e}}$ case, the roles of 
particles and anti-particles 
are reversed.  One consequence of this is
that the signs of all the other asymmetries are also reversed. 
The quantity $\delta_1 Y_P$ will obviously be significantly affected by 
this ambiguity in sign, while $\delta_2 Y_P$ will not be 
affected at all. This means that we have two possible values 
for the overall change in the effective number of neutrino 
flavours during BBN. The results of the numerical
work is presented in Figs.1-3.  
In Figure 1 we show the evolution of $L_{\nu_e}$ for
three examples, $\delta m^2_{es}/eV^2 = -0.1, -1, -10$.
We emphasise that the evolution is approximately independent
of $\sin^2 2\theta_{es}$ so long as 
$\sin^2 2\theta_{es}$ satisfies
\begin{equation}
{\rm few}\times 10^{-10} \stackrel{<}{\sim}
\sin^2 2\theta_{es} \stackrel{<}{\sim}
{\rm few}\times 10^{-5}\left( {eV^2 \over 
|\delta m^2|}\right)^{1 \over 2},
\label{rau}
\end{equation}
where the lower bound comes from adiabaticity while 
the upper bound comes from the requirement
that $\nu_e \leftrightarrow \nu_s$ oscillations
do not populate the sterile states at high temperatures
before the neutrino asymmetry is initially generated,
i.e. for $T \stackrel{>}{\sim} T_c$ 
(see e.g. Ref.\cite{fv1}).
As explained earlier, we start the low temperature
evolution at $T \sim T_c/2$ with the value of
$p_{res}/T \sim 0.5$.
Of course the full evolution from $T \gg T_c$ to
$T \to 0$ can be obtained from numerical integration
of the quantum kinetic equations (see e.g. Ref.\cite{f}).
However, for the applications considered in this paper
only the low temperature evolution is required,
which is why we use the much simpler 
formalism based on Eq.(\ref{presevol}).

The implications for BBN are shown in Figure 2,3.
Figure 2 treats the $L_{\nu_e} > 0$ case, 
while Fig.3 displays the $L_{\nu_e} < 0$ case.
As these figures show, the effect of the 
$\nu_e \leftrightarrow \nu_s$ oscillations on BBN
is very significant and depends sensitively on the
sign of the asymmetry and on the $\delta m^2_{es}$ value.
We emphasise that our equations contain
approximations. The most important are that the
repopulation is handled approximately via Eq.(\ref{dfd}).
It is obviously difficult to estimate the size of this
uncertaintly without computing repopulation
exactly. Nevertheless we expect that this 
theoretical uncertainty is typically less than 
$\delta N_{eff}^{BBN} \sim 0.2$.

It is evident that $\delta_2 Y_P$ is close to zero for 
the range of $\delta m^2_{es}$ considered.
This can be approximately understood
by noting that the generation of sterile states
occurs below the
kinetic decoupling temperature for $\nu_e$'s. This
means that the $\nu_e$ states which have converted into 
sterile states are not significantly repopulated.  
It is interesting to observe that $\delta_2 Y_P$ is computed
to be slightly negative for values of $\delta m^2_{es} \sim
{\rm few} \ eV^2$.  This feature can be qualitatively
explained as follows. As $p/T$ 
moves through the initial part of the spectrum the
MSW transitions deplete the $\overline \nu_e$'s before the
$\overline \nu_e$ spectrum is significantly distorted
by the $L_{\nu_e}$ asymmetry. However,
by the time $p/T$ moves to the higher momentum part (i.e. $p/T
\stackrel{>}{\sim} 4$), the neutrino asymmetry is so large
that the number of $\overline \nu_e$ states are significantly reduced
(c.f. with the Fermi-Dirac distribution with zero chemical
potential).
For this reason the oscillations deplete more low momentum
neutrinos than high momentum ones. The concurrent 
thermalisation
of the neutrino distribution evidently reduces the average $p/T$
per neutrino, and hence the energy density {\it divided by
$T^4$} can be slightly reduced. 
Of course the temperature of the $\overline \nu_e$ states 
would be expected to increase a little, which in 
our approximation is neglected
(much of this temperature increase would be absorbed by the
other flavours which are still in approximate thermal
equilibrium down to temperatures 1-2 MeV).

Note that in the case of very small $|\delta m^2_{es}|/eV^2
\sim 10^{-2}$, the dominant effect of the neutrino asymmetry
in the $L_{\nu_e} > 0$ case
is the modification to the Pauli blocking of the neutron decay.
Ordinarily neutrino asymmetries lead to a negligible 
effect for neutron decay however in this case 
the effect is small (but not completely negligible)
because the neutrino asymmetry is so large ($L_{\nu_e}
\simeq 0.37$ for $\delta m^2_{es}/eV^2 \sim -10^{-2}$).

Finally note that our results are quite different to
the results of Ref.\cite{sfa}. 
We do not know why this is the case. Unfortunately,
Ref.\cite{sfa} gives few details of how they
computed the evolution of the neutrino asymmetry.
Thus it is difficult to know whether the difference
lies in the details of the asymmetry evolution or
in the BBN code.

This concludes our study of the direct production
of $L_{\nu_e}$ from $\nu_e \leftrightarrow \nu_s$ oscillations.
We now consider the alternative, but more 
complicated case of the indirect production
of $L_{\nu_e}$ from a large $L_{\nu_\tau}$ asymmetry.

\section{Case 2: Implications for BBN in the
four neutrino scenario with $m_{\nu_\tau} \gg m_{\nu_\mu}, m_{\nu_e},
m_{\nu_s}$}

We now discuss the second scenario of Ref.\cite{fv2}, that is
a four neutrino model with
\begin{equation}
m_{\nu_\tau} \gg m_{\nu_\mu}, m_{\nu_e}, m_{\nu_s}
\label{ppq}
\end{equation}
We will first deviate from Ref.\cite{fv2} by considering
the simpler case where the muon neutrino is ignored.
Strictly, this would only be possible if the mixing
angle satisfies $\sin^2 2\theta_{\mu \tau} \stackrel{<}{\sim}
{\rm few} \times 10^{-10}$. 
This case is simpler because there are then only two
important oscillation modes, $\nu_\tau \leftrightarrow
\nu_s$ and $\nu_\tau \leftrightarrow \nu_e$.
Moreover these two oscillation modes always have quite different
resonance momentum, so that they can each be described by
two flavour oscillations.  Later (in part B) we will consider the
alternative case where $\sin^2 2\theta_{\mu \tau} \stackrel{>}{\sim}
{\rm few} \times 10^{-10}$ and $\nu_{\tau} \leftrightarrow
\nu_\mu$ oscillations are also important.

\subsection{Decoupled muon neutrino}

If we ignore the muon neutrino then
there are two oscillation modes with approximately the
same $|\delta m^2|$, which we denote as 
$\delta m^2_{\text{large}}$:
\begin{equation}
\nu_\tau \leftrightarrow  \nu_s, \quad \nu_\tau \leftrightarrow 
\nu_e. 
\label{3}
\end{equation}
Note that $\delta m^2_{\text{large}} \simeq m^2_{\nu_{\tau}}$
given Eq.(\ref{ppq}). 
We will consider the parameter space region where the $\delta
m^2$ values of all the other oscillation modes are small enough so that
they can be approximately neglected for temperatures
$T \stackrel{>}{\sim} 0.4 \ \text{MeV}$.
(this will be true if the $\delta m^2$ of these
other oscillation modes are all
much less than about $1 eV^2$).
This last condition means that these modes will
not significantly affect the neutron/proton ratio and hence 
cannot significantly affect BBN.

In the following discussion we consider the
case $L_{\nu_\tau} > 0$ for definiteness. This means that
the $\overline{\nu}_\tau \leftrightarrow \overline{\nu}_s$ 
generate $L_{\nu_\tau}$ while $\overline{\nu}_\tau
\leftrightarrow \overline{\nu}_e$ oscillations
reprocess some of this asymmetry into $L_{\nu_e}$.  
Note that in this scenario the sign of $L_{\nu_e}$
is necessarily the same as the sign of $L_{\nu_\tau}$.

The evolution of this system
can be described by a straightforward generalisation
of the two-flavour case given in Eq.(\ref{presevol}). 
In this case there are two MSW resonances,
$\overline \nu_\tau \leftrightarrow \overline \nu_s$ 
and $\overline \nu_\tau \leftrightarrow \overline \nu_e$.
We denote the resonance momentum of these two oscillations
by $P_1$ and $P_2$ respectively.
They are related to the neutrino asymmetries and temperature
through the equations,
\begin{equation}
{P_1 \over T} = {\delta m^2_{\text{large}} \over 
a_0 T^4 L_1}, \
{P_2 \over T} = {\delta m^2_{\text{large}} \over 
a_0 T^4 L_2},
\label{77}
\end{equation}
where $a_0 \equiv 4\sqrt{2}\zeta(3)G_F/\pi^2$ and\footnote{
We neglect the $L_{\nu_\mu}$ asymmetry (and small 
baryon/electron asymmetries) which is a valid thing to do 
provided that it is less than about $10^{-5}$.}
\begin{equation}
L_1 \equiv  
2L_{\nu_\tau} + L_{\nu_e},\
L_2 \equiv
L_{\nu_\tau} - L_{\nu_e}.
\label{99}
\end{equation}
The evolution of the lepton numbers can be obtained
by a straightforward generalisation to Eqs.(\ref{he}, 
\ref{he6}, \ref{presderiv}, \ref{presevol}),
\begin{eqnarray}
{dL_{\nu_\tau}  \over dT} &=&
-X_1 \left|{d(P_1/T)\over dT}\right| - 
X_2\left|{d(P_2/T) \over dT}\right|, \nonumber \\
{dL_{\nu_e} \over dT} &=& 
X_2\left|{d(P_2/T) \over dT}\right|, \nonumber \\
\ {dL_{\nu_\mu} \over dT} &=& 0,
\label{1az}
\end{eqnarray}
where
\begin{eqnarray}
X_1 \equiv {T \over n_\gamma}\left[
N_{\overline \nu_\tau}(P_1) -
N_{\overline \nu_s}(P_1) 
\right], \nonumber \\
X_2 \equiv {T \over n_\gamma}\left[
N_{\overline \nu_\tau} (P_2) -
N_{\overline \nu_e}(P_2) \right]. 
\label{2az}
\end{eqnarray}
Expanding out Eq.(\ref{1az}) we find
\begin{eqnarray}
y_1 {dL_{\nu_\tau} \over dT} &=& \alpha + \beta {dL_{\nu_e} \over dT},
\nonumber \\
y_2 {dL_{\nu_e} \over dT} &=& \delta + \rho {dL_{\nu_\tau} 
\over dT} ,
\label{3az}
\end{eqnarray}
where
\begin{eqnarray}
y_1 &\equiv& 1 - f_1X_1{\partial(P_1/T) \over \partial L_{\nu_\tau}}
- f_2X_2 {\partial(P_2/T) \over \partial L_{\nu_\tau}}
= 1 + {2f_1X_1 P_1 \over TL_1} + 
{f_2X_2P_2 \over TL_2}, \nonumber \\
y_2 &\equiv& 1 + f_2X_2{\partial(P_2/T) 
\over \partial L_{\nu_e}}
= 1 + {f_2X_2 P_2 \over TL_2}, \nonumber \\
\alpha &\equiv& f_1X_1{\partial (P_1/T) \over \partial T} 
+ f_2X_2{\partial (P_2/T) \over \partial T} = -4f_1X_1P_1/T^2
- 4f_2X_2P_2/T^2,
\nonumber \\ 
\beta &\equiv& f_1X_1 {\partial (P_1/T) \over \partial L_{\nu_e}}
+ f_2X_2 {\partial (P_2/T) \over \partial L_{\nu_e}} = 
{-f_1X_1 P_1 \over TL_1} + {f_2X_2 P_2 \over TL_2},
\nonumber \\
\delta &\equiv& -f_2X_2 {\partial (P_2/T) \over \partial T} 
= 4f_2X_2P_2/T^2,
\nonumber \\
\rho &\equiv& -f_2X_2{\partial (P_2/T) \over \partial 
L_{\nu_\tau}} =
{f_2X_2P_2 \over TL_2} , \nonumber \\
\label{4a}
\end{eqnarray}
and $f_i = 1$ for $d(P_i/T)/dt > 0$ and $f_i = -1$
for $d(P_i/T)/dt < 0$ ($i=1,2$).
Solving Eq.(\ref{3az}) we find,
\begin{eqnarray}
{dL_{\nu_\tau} \over dT} &=& {\delta \beta  
+ y_2 \alpha  \over y_2 y_1  - \rho \beta },
\nonumber \\
{dL_{\nu_e} \over dT} &=& {\delta +\rho {dL_{\nu_\tau} \over dT}\over
y_2 }.
\label{5az}
\end{eqnarray}
In order to integrate these equations we
need to specify the values of $L_{\nu_\alpha}$
(or equivalently $P_i/T$) at $T=T_c/2$\footnote{
Of course we also need to specify the initial signs, $f_i$,
which we take as positive. Subsequent evolution does not change
these signs.}. 
The high temperature evolution typically does
note generate significant $L_{\nu_e}$,
i.e. typically $L_{\nu_e} \ll L_{\nu_\tau}$.
So that we have $P_1/T \sim 0.5, \ P_2/T \simeq 2P_1/T$.

The evolution of the number densities is 
treated in a similar fashion to the previous section.
Specifically, the MSW transitions effect the adiabatic conversions
\begin{equation}
|\overline \nu_\tau \rangle \leftrightarrow 
|\overline \nu_s\rangle, \
|\overline \nu_\tau \rangle \leftrightarrow 
|\overline \nu_e\rangle, 
\end{equation}
at $p = P_1$ and $p=P_2$ respectively.
This means that as $P_1$, $P_2$ sweeps through 
the $\overline \nu_\tau, \ \overline \nu_e$ momentum distributions, 
\begin{eqnarray}
N_{\overline \nu_s}(P_1) &\to& N_{\overline \nu_\tau}(P_1), \nonumber \\
N_{\overline \nu_\tau}(P_1) &\to& N_{\overline \nu_s}(P_1),\nonumber \\
N_{\overline \nu_e}(P_2) &\to& N_{\overline \nu_\tau}(P_2), \nonumber \\
N_{\overline \nu_\tau}(P_2) &\to& N_{\overline \nu_e}(P_2).
\label{xp1}
\end{eqnarray}
In our numerical work the continuous momentum distribution for each flavour
is replaced by a finite number of `cells' on a logarithmically
spaced mesh. As the momentum $P_1$, $P_2$ passes
a cell, the number density in the cell
is modified according to Eq.(\ref{xp1}).
Of course weak interactions will repopulate
some of these cells as they thermalise the
neutrino momentum distributions.
We take re-population into account approximately with rate equations
of the form Eq.(\ref{dfd}) (for both $\nu_\alpha = \nu_e$
and $\nu_\alpha = \nu_\tau$) and compute the chemical potentials
via the procedure discussed in the appendix.

As in the case of the previous section, the evolution is
approximately independent of $\sin^2 2\theta_{\tau s}, 
\ \sin^2 2\theta_{\tau e}$ so long as 
\begin{eqnarray}
{\rm few}\times 10^{-10} \stackrel{<}{\sim}
&\sin^2 2\theta_{\tau s}&
 \stackrel{<}{\sim}
{\rm few}\times 10^{-5}\left( {eV^2 \over 
\delta m^2_{\text{large}}}\right)^{1 \over 2},
\nonumber \\
{\rm few}\times 10^{-10} \stackrel{<}{\sim}
&\sin^2 2\theta_{\tau e}&,
\end{eqnarray}
where the lower bounds comes from adiabaticity while 
the upper bound comes from the requirement
that $\nu_\tau \leftrightarrow \nu_s$ oscillations
do not populate the sterile states at high temperatures
before the neutrino asymmetry is initially generated.
Of course we have implicitly assumed that the
$\nu_e \leftrightarrow \nu_s,\ \nu_\mu \leftrightarrow \nu_s$
oscillation modes do not significantly populate the
sterile neutrinos before BBN.   
This is a valid assumption for a large 
range of parmaters even if
the $\nu_s$ mixes with large mixing angles 
with $\nu_{\mu}$ and/or $\nu_e$.
For example, if $\nu_\mu \leftrightarrow
\nu_s$ oscillations are approximately maximal with
$|\delta m^2_{\mu s}| \simeq 3\times 10^{-3}\ eV^2$ 
(as suggested by the atmospheric neutrino anomaly)
then the $\nu_\mu \leftrightarrow \nu_s$ oscillations
can potentially populate the sterile neutrinos at
the high temperature $T \approx 8\ MeV$.
However this doesn't happen if the $\nu_\tau$ is
in the eV mass range (for a large range of 
$\sin^2 2\theta_{\tau s}$)\cite{fv1,f}.

In Figure 4 we plot the evolution of the
neutrino asymmetries, $L_{\nu_\tau}, L_{\nu_e}$, for
three examples, $\delta m^2_{\text{large}}/eV^2 = 0.1, 10, 
1000$.  
As these figures show, for low values of
$\delta m^2_{\text{large}}/eV^2$ the 
transfer of $L_{\nu_e}$ from $L_{\nu_\tau}$ is
not very efficient. This is expected because
the transfer of asymmetry relies on repopulation
to distribute the $L_{\nu_\tau}$ away from
the momentum region
where it is created (i.e. $p \simeq P_1$ ) 
to the momentum region where
$\overline \nu_\tau
\leftrightarrow \overline \nu_e$ oscillations
are important (i.e. at $p \sim P_2$).
As $\delta m^2_{\text{large}}$ increases
the temperature where the $L_{\nu_\tau}$
asymmetry is created increases which makes the
transfer to $L_{\nu_e}$ more efficient because
of the faster repopulation rate.

The implications for BBN are shown in Figures 5,6.
Figure 5 treats the $L_{\nu_e} > 0$ case, 
while Fig.6 displays the $L_{\nu_e} < 0$ case.
As before, our results have a theoretical uncertainty 
which is dominated by the approximate treatment of repopulation.
This uncertainty is expected to be typically less than
about $\delta N_{eff}^{BBN} \sim 0.2$ (obviously
we expect this uncertainty to be much smaller
than this when $\delta_1 N^{BBN}_{eff}$ is small). 
In the $L_{\nu_\tau} > 0$ case there is a dip at around
$\delta m^2 \sim -15 \ eV^2$. This can be qualitatively understood
as follows. For $\delta m^2 \sim -15\ eV^2$ the repopulation
rate is not so rapid. This has two obvious effects:
First, the thermalisation of the $\nu_\tau$ momentum distribution
is not so rapid and this would lead to less
efficient production of $L_{\nu_e}$. This effect would lead
to a decrease in $L_{\nu_e}$ (and hence decrease 
$|\delta N^{BBN}_{eff}|$).
Second, the effect of a slow repopulation rate on the $\nu_e$
distribution would be expected to have the opposite effect.
The reason is that the momentum at which $L_{\nu_e}$
creation is most significant is in the high momentum tail.
This is because $P_2/T \sim 2P_1/T$ and significant
$L_{\nu_e}$ is not generated until 
$L_{\nu_\tau}$ is sufficiently large, i.e. roughly
$P_1/T \stackrel{>}{\sim} 2$.
The distortion of the $\nu_e$ distribution in the high
momentum tail greatly enhances the effects for
BBN because these effects depend quadratically on
the neutrino momentum. Evidently our numerical work
indicates that the second effect dominates over the
first effect.

Finally, we now consider the more interesting, but
more complicated case with the muon neutrino included.

\subsection{Including the muon neutrino.}

When the muon neutrino is included (i.e.
$\sin^2 2\theta_{\mu \tau} \stackrel{>}{\sim} {\rm few}
\times 10^{-10}$) there are
three oscillation modes with approximately the
same $|\delta m^2|$, which we again denote as 
$\delta m^2_{\text{large}}$:
\begin{eqnarray}
\nu_\tau &\leftrightarrow & \nu_s, \quad \nu_\tau \leftrightarrow 
\nu_\mu, \quad \nu_\tau \leftrightarrow \nu_e.
\label{32}
\end{eqnarray}
All the other oscillation modes have much
smaller $|\delta m^2|$ values.
As before, we will consider the parameter 
space region where the $\delta
m^2$ values of all the other oscillation modes are small enough so that
they can be approximately neglected for temperatures
$T \stackrel{>}{\sim} 0.4 \ \text{MeV}$.
This last condition means that these modes will
not affect the neutron/proton ratio and hence 
cannot significantly affect BBN.

In the following discussion we again consider the
case $L_{\nu_\tau} > 0$ for definiteness. This means that
the $\overline{\nu}_\tau \leftrightarrow \overline{\nu}_s$ 
generate $L_{\nu_\tau}$ while the
other two oscillation modes reprocess some of this asymmetry
into $L_{\nu_e}, L_{\nu_\mu}$.
In Ref.\cite{fv2} this system was 
first considered in this context.
There, it was assumed that $L_{\nu_e} = L_{\nu_\mu}$.
While this is a good approximation for large enough
values of $|\delta m^2|$, it is not always valid (as
we will show below).
In the following we will not assume this and consider
the effect of the three oscillations modes.

At this point one may legitimately worry about 3-flavour
effects. This is because the resonance momentum of the
$\nu_\tau \leftrightarrow \nu_e$ and
$\nu_\tau \leftrightarrow \nu_\mu$ oscillation modes
are expected to be approximately equal.
However it turns out that these oscillations
tend to be dynamically driven apart as we will explain
latter on. Thus, it turns out that it is
actually reasonable to treat all three oscillation
modes independently as 2-flavour MSW transitions.
In the earlier paper\cite{fv2} this issue was 
not fully discussed
so our treatment here improves on Ref.\cite{fv2}.

In this system there are three MSW resonances,
$\overline \nu_\tau \leftrightarrow \overline \nu_s$, 
$\overline \nu_\tau \leftrightarrow \overline \nu_\mu$ and
$\overline \nu_\tau \leftrightarrow \overline \nu_e$.
We denote the resonance momentum of these three oscillations
by $P_1, \ P_2$ and $P_3$ respectively.
They are related to the neutrino asymmetries and temperature
through the equations,
\begin{equation}
{P_i \over T} = {\delta m^2_{\text{large}} \over 
a_0 T^4 L_i},
\label{77x}
\end{equation}
where $i=1,2,3$, $a_0 \equiv 4\sqrt{2}\zeta(3)G_F/\pi^2$ and
\begin{equation}
L_1 \equiv  
2L_{\nu_\tau} + L_{\nu_\mu} + L_{\nu_e},\
L_2 \equiv
L_{\nu_\tau} - L_{\nu_\mu},\
L_3 \equiv
L_{\nu_\tau} - L_{\nu_e}.
\label{99x}
\end{equation}
Using a simpler procedure to the previous (sub)sections,
we have
\begin{eqnarray}
{dL_{\nu_\tau}  \over dT} &=&
-X_1 \left|{d(P_1/T)\over dT}\right| - 
X_2\left|{d(P_2/T) \over dT}\right| -
X_3\left|{d(P_3/T) \over dT}\right|, \nonumber \\
{dL_{\nu_e} \over dT} &=& 
X_3\left|{d(P_3/T) \over dT}\right|, \nonumber \\
{dL_{\nu_\mu} \over dT} &=& 
X_2\left|{d(P_2/T) \over dT}\right|, 
\label{1aa}
\end{eqnarray}
where
\begin{eqnarray}
X_1 \equiv {T \over n_\gamma}\left[
N_{\overline \nu_\tau}(P_1) -
N_{\overline \nu_s}(P_1) 
\right], \nonumber \\
X_2 \equiv {T \over n_\gamma}\left[
N_{\overline \nu_\tau}(P_2) - N_{\overline \nu_\mu}(P_2) 
\right], \nonumber \\
X_3 \equiv {T \over n_\gamma}\left[
N_{\overline \nu_\tau}(P_3) - N_{\overline \nu_e}(P_3) 
\right].
\label{2aa}
\end{eqnarray}
Expanding out Eq.(\ref{1aa}) we find
\begin{eqnarray}
y_1 {dL_{\nu_\tau} \over dT} &=& \alpha + \beta {dL_{\nu_e} \over dT} + 
\gamma {dL_{\nu_\mu} \over dT},
\nonumber \\
y_2 {dL_{\nu_e} \over dT} &=& \delta + \rho {dL_{\nu_\tau} 
\over dT},
\nonumber \\
y_3 {dL_{\nu_\mu} \over dT} &=& \eta + 
\phi {dL_{\nu_\tau} \over dT},
\label{3aa}
\end{eqnarray}
where
\begin{eqnarray}
y_1 &\equiv& 1 - f_1X_1{\partial(P_1/T) \over \partial L_{\nu_\tau}}
- f_2X_2 {\partial(P_2/T) \over \partial L_{\nu_\tau}}
- f_3X_3 {\partial(P_3/T) \over \partial L_{\nu_\tau}}\nonumber \\
&=& 1 + {2f_1X_1 P_1 \over TL_1} + 
{f_2X_2P_2 \over TL_2} +
{f_3X_3P_3 \over TL_3}, \nonumber \\
y_2 &\equiv& 1 + f_3X_3{\partial(P_3/T) 
\over \partial L_{\nu_e}} =
1 + {f_3X_3 P_3 \over TL_3}, \nonumber \\
y_3 &\equiv& 1 + f_2X_2{\partial (P_2/T) \over \partial L_{\nu_\mu}}
= 1 + {f_2X_2 P_2 \over TL_2},
\nonumber \\
\alpha &\equiv& f_1X_1{\partial (P_1/T) \over \partial T} 
+ f_2X_2{\partial (P_2/T) \over \partial T} 
+ f_3X_3{\partial (P_3/T) \over \partial T} \nonumber \\
&=& -4f_1X_1P_1/T^2
- 4f_2X_2P_2/T^2 - 4f_3X_3P_3/T^2,
\nonumber \\ 
\beta &\equiv& f_1X_1 {\partial (P_1/T) \over \partial L_{\nu_e}}
+ f_3X_3 {\partial (P_3/T) \over \partial L_{\nu_e}} = 
{-f_1X_1 P_1 \over TL_1} + {f_3X_3 P_3 \over TL_3},
\nonumber \\
\gamma &\equiv& f_1X_1{\partial (P_1/T) 
\over \partial L_{\nu_\mu}} +
f_2X_2{\partial (P_2/T) 
\over \partial L_{\nu_\mu}} = 
{-f_1X_1 P_1 \over TL_1} +
{f_2X_2 P_2 \over TL_2}, 
\nonumber \\
\delta &\equiv& -f_3X_3 {\partial (P_3/T) \over \partial T} 
= 4f_3X_3P_3/T^2,
\nonumber \\
\rho &\equiv& -f_3X_3{\partial (P_3/T) \over \partial 
L_{\nu_\tau}} =
{f_3X_3P_3 \over TL_3}, \nonumber \\
\eta &\equiv& -f_2X_2{\partial (P_2/T) \over \partial T}
= 4f_2X_2 P_2/T^2,
\nonumber \\
\phi &\equiv& -f_2X_2 {\partial (P_2/T) \over \partial L_{\nu_\tau}} 
= {f_2X_2P_2 \over TL_2},
\label{4aa}
\end{eqnarray}
and $f_i = 1$ for $d(P_i/T)/dt > 0$ and $f_i = -1$
for $d(P_i/T)/dt < 0$ ($i=1,2,3$).
Solving Eq.(\ref{3aa}) we find,
\begin{eqnarray}
{dL_{\nu_e} \over dT} &=& {\delta y_3(y_1 y_3 - \gamma \phi)
+ \rho y_3(\alpha y_3 + \gamma \eta) \over
y_2 y_3(y_1 y_3 - \gamma \phi) - 
\rho \beta y_3^2},
\nonumber \\
{dL_{\nu_\tau} \over dT} &=& {\alpha y_3 + \gamma \eta +
\beta y_3{dL_{\nu_e} \over dT}\over
y_1 y_3 - \gamma \phi},
\nonumber \\
{dL_{\nu_\mu} \over dT} &=& {1 \over y_3}\left[
\eta  + \phi {dL_{\nu_\tau}
\over dT}\right].
\label{5aa}
\end{eqnarray}
In order to integrate these equations we
need to specify the values of $L_{\nu_\alpha}$
(or equivalently $P_i/T$) at $T=T_c/2$. 
The high temperature evolution typically does
not generate significant $L_{\nu_e}, L_{\nu_\mu}$
(i.e. typically $L_{\nu_e}, L_{\nu_\mu} \ll L_{\nu_\tau}$).
So that we have
$P_1/T \sim 0.5, \ P_2/T \simeq  P_3/T \simeq 2P_1/T$.
Now, there is no reason why $P_2/T$ should exactly
coincide with $P_3/T$ (although it will be approximately
equal). In fact for these two oscillations,
$P_2/T = P_3/T$ is not dynamically stable as we shall 
now explain.  The behaviour of oscillations such as these
(in the context of a quite different model)
has been studied in some detail in Ref.\cite{fv4}.
Generically there are two possible outcomes.
Either the evolution of lepton number is such that
it drives $P_2/T \to P_3/T$ {\it or}
the evolution of lepton numbers is such
as to drive them apart.
To figure out whats going to happen
imagine that $P_2/T$ is slightly less
than $P_3/T$. In other words
the $\overline \nu_\tau \leftrightarrow
\overline \nu_e$ oscillation resonance preceeds the
$\overline \nu_\tau \leftrightarrow \overline \nu_\mu$ oscillation
resonance.  This means that the $\overline \nu_\tau 
\leftrightarrow \overline \nu_e$ resonance
will efficiently interchange $\overline \nu_\tau$ 
and $\overline \nu_e$ states
at the resonance. This will transfer some
$L_{\nu_\tau}$ to $L_{\nu_e}$ and will thus
speed up the resonance a bit since it is inversely 
proportional to the difference of $L_{\nu_\tau}$ and 
$L_{\nu_e}$.  The trailing $\overline \nu_\tau
\leftrightarrow \overline \nu_\mu$
resonance will be less effective in transferring
$L_{\nu_\tau}$ to $L_{\nu_\mu}$ because
at this resonance there will be approximately
equal number of $\overline \nu_\mu $ and $\overline \nu_\tau $
states thanks to the efforts of the $\overline \nu_\tau \leftrightarrow
\overline \nu_e$ resonance.
Thus the two resonances will slowly move apart
until eventually they will be far enough apart
so that the thermalisation due to the collisions
will be rapid enough to thermalise the $\nu_\tau$
spectrum such that the $L_{\nu_\mu}$ is created
at approximately the same rate as the $L_{\nu_e}$.
For definiteness, we will assume as our initial
condition that $P_3/T > P_2/T$. 
For our numerical work we will assume that
$P_3/T = 1.01P_2/T $ initially.
This could be due
to a slightly larger $|\delta m^2_{\tau e}| >
|\delta m^2_{\tau \mu}|$ for example.
We found very similar results for even smaller choices such
as $P_3/T = 1.001 P_2/T$. 

It is straightforward to numerically
integrate the evolution equations, Eq.(\ref{5aa}), 
with $T=T_c/2$ `initial conditions' as described above.
We keep track of the number distributions of all 
the 4 flavours (using a completely analogous procedure to the
previous cases). 
In Figure 7 we plot the evolution of the
neutrino asymmetries, $L_{\nu_\tau}, L_{\nu_\mu}, L_{\nu_e}$, for
three examples, $\delta m^2_{\text{large}}/eV^2 = 0.1, 10, 
1000$.  
The implications for BBN are shown in Figures 8,9.
Figure 8 treats the $L_{\nu_e} > 0$ case, 
while Fig.9 displays the $L_{\nu_e} < 0$ case.
As these figures show, this case is very similar to the previous
results were the muon neutrino was neglected.
Of course if we had started with $P_2/T > P_3/T$ then
the $L_{\nu_\mu}$ and $L_{\nu_e}$ are approximately interchanged.
In this case, the modification to $N_{eff}^{BBN}$ would be
somewhat smaller, especially for lower values of
$\delta m^2_{\text{large}}$.

Let us compare our results with the original work in Ref.\cite{fv2}
and the more recent work of Ref.\cite{sfa}.
In Ref.\cite{fv2} we
made the approximation that the repopulation
was instantaneous above about 1.5 MeV.
We also assumed that $L_{\nu_e} = L_{\nu_\mu}$ and 
derived evolution equations consistent with that assumption.
We found that
$\delta N^{BBN}_{eff} \approx -0.5$ for $L_{\nu_e} > 0$
and
$\delta N^{BBN}_{eff} \approx 0.4$ for $L_{\nu_e} < 0$ for
\begin{equation}
10 \stackrel{<}{\sim}
\delta m^2_{\text{large}}/eV^2 \stackrel{<}{\sim} 1000.
\label{ik8}
\end{equation}
For the $L_{\nu_e} > 0$ case the results are in rough agreement
for $\delta m^2_{\text{large}} \stackrel{<}{\sim} 100 \ eV^2$.
The difference for larger $\delta m^2_{\text{large}}$ is
due to the more accurate treatment of chemical decoupling
which suggests a lower decoupling temperature.
For the
$L_{\nu_e} < 0$ case the effect is underestimated by about $0.4$
in $\delta N^{BBN}_{eff}$ in Ref.\cite{fv2}. 
This difference is partly due to a mistake in the numerical
work of Ref.\cite{fv2} which we have recently discovered.

In Ref.\cite{sfa} they consider the case of part A of this section,
i.e. neglecting the muon neutrino. Their results seems do not
seem to be consistent with ours, expecially for $L_{\nu_e} > 0$. 
We do not know the reason for this.

\section{Conclusion}

We have made a detailed study of several `four neutrino scenarios'
which can generate significant $L_{\nu_e}$ asymmetry
thereby affecting BBN (these scenarios were 
first discussed in this context in Ref.\cite{fv2}).
In the first case we considered the direct production of 
$L_{\nu_e}$ from $\nu_e \leftrightarrow \nu_s$ oscillations.
Our results are shown in Figure 2,3. Clearly very large
modifications to BBN are possible and depend
sensitively on $\delta m^2$ [but
are approximately independent of $\sin^2 2\theta$
so long as $\sin^2 2\theta$ is in the range, Eq.(\ref{rau})].
The results also depends critically on the sign of $L_{\nu_e}$.
We also studied the indirect production of $L_{\nu_e}$ from
$L_{\nu_\tau}$. Our results are given in figures 5,6
for the case where the muon neutrino can be 
neglected and figures 8,9 where the
muon neutrino is included. 
These results are in rough agreement
with our earlier conclusion\cite{fv2} that 
$\delta N^{BBN}_{eff} \sim -0.5$
for the case of positive $L_{\nu_e}$ for the parameter range
Eq.(\ref{ik8}).
Notice that the figures show a slightly larger
effect for $\delta m^2 \sim 10\ eV^2$ where
the slow repopulation rate becomes important
(Ref.\cite{fv2} assumed that repopulation was instantaneous).
Also the more accurate treatment of repopulation
in the kinetic decoupling region suggests a
larger $\delta N^{BBN}_{eff}$ for 
$|\delta m^2|/eV^2 \stackrel{>}{\sim} 100$ than
was found previously.

We conclude by emphasising once more that the detailed
predictions of models with light sterile neutrinos are
quite model dependent. Quantitatively different results occur
for four neutrino models with approximately degenerate
$\nu_\mu, \ \nu_\tau$\cite{bfv}, as well as in
six neutrino models with three light sterile neutrinos
approximately maximally mixed with each
of the ordinary neutrinos\cite{fv4}. 
It is a remarkable prospect that accurate determinations
of the primordial element abundances, as well
as forthcoming precision measurements of the
anisotropy of the cosmic microwave background may 
one day help to distinguish between competing models of
particle physics.

\appendix

\section{Repopulation}

Consider for definiteness the case of
$\overline \nu_\alpha \leftrightarrow \overline \nu_s$ 
oscillations which generate a large $L_{\nu_\alpha}$.
The $\overline \nu_\alpha \leftrightarrow \overline \nu_s$
oscillations depelete the $\overline \nu_\alpha$ states
at the MSW resonance. Elastic collisions will tend to
thermalise the momentum distributions so that
they can be approximately described by chemical 
potentials, while inelastic collisions will create and modify
the chemical potentials.
Let us denote the total elastic and inelastic collision rates
by the notation, $\Gamma^{E}_{\alpha}, \ \Gamma^I_{\alpha}$ 
respectively.  It happens that $\Gamma^E_{\alpha} 
\gg \Gamma^I_{\alpha}$\footnote{ 
Numerically $\Gamma^E_e/\Gamma^I_e \simeq 6.3$,
$\Gamma^E_{\mu,\tau}/\Gamma^I_{\mu,\tau} \simeq 8.0$.}
so it makes sense to describe the neutrino distributions
in terms of chemical potentials and a common
temperature throughout the chemical decoupling period
($2 \stackrel{<}{\sim} T/MeV \stackrel{<}{\sim} 4$).
{\it We emphasise that the actual momentum distribution 
is always computed from Eq.(\ref{dfd}), the purpose of 
this present discussion is to work out 
the evolution of the chemical potentials
which are needed in the right-hand side of Eq.(\ref{dfd}).} 

The chemical potentials are related to
the lepton number by the equation:
\begin{equation}
L_{\nu_{\alpha}} = {1 \over 4\zeta (3)}\int^{\infty}_0
{x^2 dx \over 1 + e^{x + \tilde{\mu}_{\alpha}}} - 
{1 \over 4\zeta (3)}\int^{\infty}_0
{x^2 dx \over 1 + e^{x + \tilde{\mu}_{\overline{\alpha}}}},  
\end{equation}
where $\tilde{\mu}_{\alpha} \equiv \mu_{\nu_\alpha}/T$
and $\tilde{\mu}_{\overline{\alpha}} \equiv 
\mu_{\overline{\nu}_{\alpha}}/T$ and
$\zeta (3) \simeq 1.202$ is the Riemann zeta function of 3.
Expanding out the above equation,
\begin{equation}
L_{\nu_{\alpha}} \simeq -{1 \over 24\zeta (3)}\left[
\pi^2 (\tilde{\mu}_{\alpha} - 
\tilde{\mu}_{\overline{\alpha}})
- 6(\tilde{\mu}_{\alpha}^2 - 
\tilde{\mu}_{\overline{\alpha}}^2)\ln 2
+ (\tilde{\mu}_{\alpha}^3 - 
\tilde{\mu}_{\overline{\alpha}}^3) \right].
\label{j1}
\end{equation}
This is an exact equation for $\tilde{\mu}_{\alpha} =
- \tilde{\mu}_{\overline{\alpha}}$, otherwise it holds to a good
approximation provided that $|\tilde{\mu}_{\alpha,\overline{\alpha}}|
\stackrel{<}{\sim} 1$. 
If we turn off inelastic collisions for a moment,
and assume that the oscillation
generated $L_{\nu_\alpha}$ is positive then
the oscillations generate a large $\tilde{\mu}_{\overline{\alpha}}$.
The evolution of $\tilde{\mu}_{\overline{\alpha}}$
due to the generation of $L_{\nu_\alpha}$ (i.e.
due to oscillations) can easily be obtained from Eq.(\ref{j1}),
\begin{equation}
\left.
{d\tilde{\mu}_{\overline{\alpha}}\over dT}\right|_{osc} \simeq 
\left[ {24\zeta(3) \over
\pi^2 - 12\tilde{\mu}_{\overline {\alpha}}ln 2 + 
3\tilde{\mu}^2_{\overline {\alpha}}}\right]
{dL_{\nu_\alpha} \over dT}, \
\left.
{d\tilde{\mu}_{\alpha}\over dT}\right|_{osc} \simeq 0.
\label{ppq2}
\end{equation}
Now, let us turn on the inelastic collisions and see what happens.
There are six inelastic processes, which we list below together
with their thermally averaged interaction rates\cite{ekt}
\vskip 0.9cm
\begin{tabular}{|l|l|}\hline
{\em process}&
{\em rate}\\ \hline
(1)\
$\nu_\tau \overline \nu_\tau \leftrightarrow \nu_\mu \overline \nu_\mu$
& $\Gamma^I_1 = F_0$ \\
(2)\
$\nu_\tau \overline \nu_\tau \leftrightarrow \nu_e \overline \nu_e$
& $\Gamma^I_2 = F_0$ \\
(3)\
$\nu_\tau \overline \nu_\tau \leftrightarrow e^+ e^- $
& $\Gamma^I_3 = (8x^2 - 4x + 1)F_0$ \\
(4)\
$\nu_\mu \overline \nu_\mu \leftrightarrow \nu_e \overline \nu_e$
& $\Gamma^I_4 = F_0$ \\
(5)\
$\nu_\mu \overline \nu_\mu \leftrightarrow e^+ e^- $
& $\Gamma^I_5 = (8x^2 - 4x + 1)F_0 $\\
(6)\
$\nu_e \overline \nu_e \leftrightarrow e^+ e^- $
& $\Gamma^I_6 = (8x^2 + 4x + 1)F_0 $\\
\hline
\end{tabular}
\vskip 0.9cm
\noindent
In the above table,
$x \equiv \sin^2 \theta_w$ is the weak mixing angle
($\sin^2 \theta_w \simeq 0.23$), and $F_0 \equiv 
{G^2_F\langle p\rangle^2 \over 6\pi}n_\gamma \simeq 0.13G_F^2 T^5$.
Consider first the first process listed
in the above table.
This process will change the number of
$\nu_\tau, \overline \nu_\tau, \nu_\mu, \overline \nu_\mu$
states such that
\begin{equation}
\left.  {dn_{\nu_\tau}\over dt}\right|_{(1)} = 
\left.  {dn_{\overline \nu_\tau} \over dt}\right|_{(1)} = 
-\left.  {dn_{\nu_\mu} \over dt}\right|_{(1)} = 
-\left.  {dn_{\overline \nu_\mu} \over dt}\right|_{(1)}, 
\end{equation}
where the subscript `$\left. \right|_{(1)}$' denotes
the contribution to the rate of change due to the process
(1) in the table. 
The rate can be expressed approximately as follows,
\begin{equation}
\left.
{d(n_{\nu_\tau}/n_0) \over dt}\right|_{(1)} = 
{1 \over n_0}
\int_0^{\infty} \int_0^{\infty} \left[
N_{\nu_\mu}(p) N_{\overline \nu_\mu}(p') -
N_{\nu_\tau}(p) N_{\overline \nu_\tau}(p') \right]
\sigma_1 (p,p') dp dp'
\label{kil}
\end{equation}
For convenience we have normalised with respect to the number of
neutrinos in a Fermi-Dirac distribution with zero chemical potential,
$n_0 = {3\zeta (3) T^3\over 4 \pi^2}$.
Let us further make the useful approximation that
\begin{equation}
N_{\nu_\alpha}(p) \simeq e^{-\tilde{\mu}_{\alpha}} N_0(p),\
N_{\overline \nu_\alpha}(p) \simeq 
e^{-\tilde{\mu}_{\overline \alpha}} N_0(p),
\ N_{e^\pm}(p) = N_0(p),
\end{equation}
where $N_0(p)$ is the Fermi-Dirac distribution with zero chemical
potential.
Note that the $e^{\pm}$ distributions have zero chemical potential
due to the very rapid collisions with the background photons
(see e.g.Ref\cite{weinberg}).
Thus, with the above approximation, Eq.(\ref{kil}) can be
expressed in the simple form,
\begin{equation}
\left. {d\eta_{\nu_\tau} \over dt}\right|_{(1)}
= \left(\eta_{\nu_\mu}\eta_{\overline \nu_\mu} -
\eta_{\nu_\tau}\eta_{\overline \nu_\tau} \right)
\Gamma_1^I ,
\end{equation}
where $\eta_{\nu_\alpha} \equiv e^{\tilde{\mu}_\alpha}$,
$\eta_{\overline \nu_\alpha} \equiv 
e^{\tilde{\mu}_{\overline \alpha}}$.
Clearly, we also have that
\begin{eqnarray}
\left.  {d\eta_{\nu_\tau} \over dt}\right|_{(1)}
&=& \left.  {d\eta_{\overline\nu_\tau} \over dt}\right|_{(1)}
= -\left.  {d\eta_{\nu_\mu} \over dt}\right|_{(1)}
= -\left.  {d\eta_{\overline\nu_\mu} \over dt}\right|_{(1)},
\nonumber \\
\left.  {d\eta_{\nu_e} \over dt}\right|_{(1)}
&=& \left.  {d\eta_{\overline\nu_e} \over dt}\right|_{(1)}
= 0.
\end{eqnarray}
A similar set of equations can be obtained for
the other five inelastic processes. Putting this
altogether, we have
\begin{eqnarray}
\left.  {d\eta_{\nu_\tau}\over dt}\right|_{repop} &=& 
\left.  {d\eta_{\overline \nu_\tau}\over dt}\right|_{repop} =
\sum_{i=1}^{6} \left.{d\eta_{\nu_\tau} \over dt}\right|_{(i)}
\nonumber \\
&=& \left( \eta_{\nu_\mu}\eta_{\overline \nu_\mu} - 
\eta_{\nu_\tau} \eta_{\overline \nu_\tau}\right)\Gamma_1^I +
\left( \eta_{\nu_e}\eta_{\overline \nu_e} - 
\eta_{\nu_\tau} \eta_{\overline \nu_\tau}\right)\Gamma_2^I +
\left(1 - \eta_{\nu_\tau}\eta_{\overline\nu_\tau}
\right)\Gamma_3^I,
\nonumber \\
\left.  {d\eta_{\nu_\mu}\over dt}\right|_{repop} &=& 
\left.  {d\eta_{\overline \nu_\mu}\over dt}\right|_{repop} =
\sum_{i=1}^{6} \left.{d\eta_{\nu_\mu} \over dt}\right|_{(i)}
\nonumber \\
&=& \left( \eta_{\nu_\tau}\eta_{\overline \nu_\tau} - 
\eta_{\nu_\mu} \eta_{\overline \nu_\mu}\right)\Gamma_1^I +
\left( \eta_{\nu_e}\eta_{\overline \nu_e} - 
\eta_{\nu_\mu} \eta_{\overline \nu_\mu}\right)\Gamma_4^I +
\left(1 - \eta_{\nu_\mu}\eta_{\overline\nu_\mu}
\right)\Gamma_5^I,
\nonumber \\
\left.  {d\eta_{\nu_e}\over dt}\right|_{repop} &=& 
\left.  {d\eta_{\overline \nu_e}\over dt}\right|_{repop} =
\sum_{i=1}^{6} \left.{d\eta_{\nu_e} \over dt}\right|_{(i)}
\nonumber \\
&=& \left( \eta_{\nu_\tau}\eta_{\overline \nu_\tau} - 
\eta_{\nu_e} \eta_{\overline \nu_e}\right)\Gamma_2^I +
\left( \eta_{\nu_\mu}\eta_{\overline \nu_\mu} - 
\eta_{\nu_e} \eta_{\overline \nu_e}\right)\Gamma_4^I +
\left(1 - \eta_{\nu_e}\eta_{\overline\nu_e}
\right)\Gamma_6^I.
\end{eqnarray}
Of course the total rate of change of $\tilde{\mu}_{\alpha},
\ \tilde{\mu}_{\overline \alpha}$ is given by
\begin{eqnarray}
{d\tilde{\mu}_{\alpha}\over dt} =
\left.{d\tilde{\mu}_{\alpha}\over dt}\right|_{osc} + 
\left.{d\tilde{\mu}_{\alpha}\over dt}\right|_{repop}, \nonumber \\ 
{d\tilde{\mu}_{\overline \alpha}\over dt} =
\left.{d\tilde{\mu}_{\overline \alpha}\over dt}\right|_{osc} + 
\left.{d\tilde{\mu}_{\overline \alpha}\over dt}\right|_{repop}.
\end{eqnarray}
The above equation can be 
used to approximately compute the set of chemical potentials.
Then using Eqs.(\ref{dfd}) the evolution of the
set of number distributions can be obtained.

In Figure 10 we give the evolution of
$\tilde{\mu}_{\alpha, \overline \alpha}$ for
some illustrative examples. In the Figures we
consider the model discussed in section IV (part A)
with the parameter choice $\delta  m^2_{\text{large}} =
10 \ eV^2$ for Figure 10a and $\delta m^2_{\text{large}} =
200 \ eV^2$ for Figure 10b.

We have also compared the above repopulation procedure 
with the simpler procedure of a fixed decoupling temperature,
$T_{dec}^{\alpha}$ where $\mu_{\nu_\alpha} = 
-\mu_{\overline \nu_\alpha}$ for $T > T_{dec}^{\alpha}$ 
and $\mu_{\nu_\alpha}$ frozen (in the case $L_{\nu_\alpha} > 
0$) for $T < T_{dec}^{\alpha}$.  We get rough agreement 
provided that $T_{dec}^{\alpha} \approx 3\ MeV$.

\acknowledgments{
The author thanks R. R. Volkas and T. L. Yoon for comments.
The author is an Australian research fellow.}

%\end{document}
\newpage
\centerline{\large {\bf Figure Captions}}
\vskip 0.5cm
\noindent
Figure 1: Low temperature evolution of $L_{\nu_e}/h$ 
($h \equiv T^2_{\nu}/T^3_{\gamma}$) due
to $\nu_e \leftrightarrow \nu_s$ oscillations, for
the parameter choices, $\delta m^2 = -0.1\ eV^2$ (solid line),
$\delta m^2 = -1\ eV^2$ (dashed line) and $\delta m^2 = -10
\ eV^2$ (dashed-dotted line).

\vskip 0.5cm
\noindent
Figure 2: $\delta N_{eff}^{BBN}$ versus $|\delta m^2_{es}|$.
The dashed line is the contribution $\delta_1 N_{eff}^{BBN}$
due to the effects of the $L_{\nu_e}$ asymmetry while
the dashed-dotted line is the contribution $\delta_2 N_{eff}^{BBN}$
due to the change in the expansion rate. The solid line
is the total contribution $\delta N_{eff}^{BBN} = 
\delta_1 N_{eff}^{BBN} + \delta_2 N_{eff}^{BBN}$. 
This figure considers the case $L_{\nu_e} > 0$.

\vskip 0.5cm
\noindent
Figure 3: Same as Figure 2 except $L_{\nu_e} < 0$ is considered.

\vskip 0.5cm
\noindent
Figure 4: Low temperature evolution of $L_{\nu_\tau}/h, \
L_{\nu_e}/h$ ($h \equiv T^3_{\nu}/T^3_{\gamma}$)
for the model of case 2 (part A).
Figs. 4a,4b,4c correspond to the parameter choices
$\delta m^2_{\text{large}} = 0.1\ eV^2$, 
$\delta m^2_{\text{large}} = 10\ eV^2$ and 
$\delta m^2_{\text{large}} = 1000 \ eV^2$ respectively.

\vskip 0.5cm
\noindent
Figure 5: $\delta N_{eff}^{BBN}$ versus 
$\delta m^2_{\text{large}}$ for the model of case 2 (part A).
The dashed line is the contribution $\delta_1 N_{eff}^{BBN}$
due to the effects of the $L_{\nu_e}$ asymmetry while
the dashed-dotted line is the contribution $\delta_2 N_{eff}^{BBN}$
due to the change in the expansion rate. The solid line
is the total contribution $\delta N_{eff}^{BBN} = 
\delta_1 N_{eff}^{BBN} + \delta_2 N_{eff}^{BBN}$. 
This figure considers the case $L_{\nu_e} > 0$.

\vskip 0.5cm
\noindent
Figure 6: Same as Figure 5 except $L_{\nu_e} < 0$ is considered.

\vskip 0.5cm
\noindent
Figure 7: Low temperature evolution of $L_{\nu_\tau}/h,\
L_{\nu_\mu}/h, \ L_{\nu_e}/h$ ($h \equiv T^3_{\nu}/T^3_{\gamma}$)
for the model of case 2 (part B).
Figs. 7a,7b,7c correspond to the parameter choices
$\delta m^2_{\text{large}} = 0.1\ eV^2$, 
$\delta m^2_{\text{large}} = 10\ eV^2$ and 
$\delta m^2_{\text{large}} = 1000 \ eV^2$ respectively.

\vskip 0.5cm
\noindent
Figure 8: $\delta N_{eff}^{BBN}$ versus 
$\delta m^2_{\text{large}}$ for the model of case 2 (part B).
The dashed line is the contribution $\delta_1 N_{eff}^{BBN}$
due to the effects of the $L_{\nu_e}$ asymmetry while
the dashed-dotted line is the contribution $\delta_2 N_{eff}^{BBN}$
due to the change in the expansion rate. The solid line
is the total contribution $\delta N_{eff}^{BBN} = 
\delta_1 N_{eff}^{BBN} + \delta_2 N_{eff}^{BBN}$. 
This figure considers the case $L_{\nu_e} > 0$.

\vskip 0.5cm
\noindent
Figure 9: Same as Figure 8 except $L_{\nu_e} < 0$ is considered.

\vskip 0.5cm
\noindent
Figure 10: Evolution of $\tilde{\mu}_{\alpha, \overline \alpha}$
for the model of case 2 (part A).  Fig 10a, 10b corresponds to the
parameter choices $\delta m^2_{\text{large}} = 10\ eV^2$ and
$\delta m^2_{\text{large}} = 200 \ eV^2$ respectively.
In the figures the thin solid, dashed, dashed-dotted, dotted
and thick solid lines correspond to $\tilde{\mu}_{e}, \
\tilde{\mu}_{\mu}, \ \tilde{\mu}_{\tau}, \ \tilde{\mu}_{\overline e},\
\tilde{\mu}_{\overline \tau}$ respectively.
Note that $\tilde{\mu}_{\overline \mu}
= \tilde{\mu}_{\mu}$ in this case.

% xxxxxxxxxxx
%\end{document}

\epsfig{file=xf1.eps,width=15cm}
\newpage
\epsfig{file=xf2.eps,width=15cm}
\newpage
\epsfig{file=xf3.eps,width=15cm}
\newpage
\epsfig{file=xf4a.eps,width=15cm}
\newpage
\epsfig{file=xf4b.eps,width=15cm}
\newpage
\epsfig{file=xf4c.eps,width=15cm}
\newpage
\epsfig{file=xf5.eps,width=15cm}
\newpage
\epsfig{file=xf6.eps,width=15cm}
\newpage
\epsfig{file=xf7a.eps,width=15cm}
\newpage
\epsfig{file=xf7b.eps,width=15cm}
\newpage
\epsfig{file=xf7c.eps,width=15cm}
\newpage
\epsfig{file=xf8.eps,width=15cm}
\newpage
\epsfig{file=xf9.eps,width=15cm}
\newpage
\epsfig{file=xf10a.eps,width=15cm}
\newpage
\epsfig{file=xf10b.eps,width=15cm}

\begin{references}

\bibitem{ftv}
R. Foot, M. J. Thomson and R. R. Volkas, 
Phys. Rev. D53, 5349 (1996).

\bibitem{fv1}
R. Foot and R. R. Volkas, Phys. Rev. D55, 5147 (1997).

\bibitem{fv2} 
R. Foot and R. R. Volkas, Phys. Rev. D56, 6653 (1997);
Erratum-ibid, D59, 029901 (1999).

\bibitem{f}
R. Foot, Astropart. Phys. 10, 253 (1999).

\bibitem{bvw}
N. F. Bell, R. R. Volkas and Y.Y.Y.Wong, Phys. Rev. D59, 113001 (1999).

\bibitem{bfv}
N. F. Bell, R. Foot and R. R. Volkas, Phys. Rev. D58, 105010 (1998).

\bibitem{fv3}
R. Foot and R. R. Volkas, Astropart. Phys. 7, 283 (1997).

\bibitem{fv4}
R. Foot and R. R. Volkas, hep-ph/9904336 (1999). 

\bibitem{sfa}
X. Shi, G. M. Fuller and K. Abazajian, hep-ph/9905259, v2.

\bibitem{bbn1}
For some recent studies,
see e.g. N. Hata et al., Phys. Rev. Lett. 75, 3977 (1995);
P. J. Kernan and S. Sarkar, Phys. Rev. D54, 3681 (1996);
K. A. Olive and D. Thomas, astro-ph/9811444;
S. Burles et al., astro-ph/9901157;
E. Lisi et al., Phys. Rev. D59, 123520 (1999).
 

\bibitem{ka}
See, for example, 
K. Kainulainen, H. Kurki-Suonio and E. Sihvola, Phys. Rev. D59, 083505
(1999).

\bibitem{weinberg}
S. Weinberg, Gravitation and Cosmology, Wiley 1972, chapter 15.

\bibitem{olive}
K. A. Olive, D. N. Schramm, D. Thomas and T. P. Walker,
Phys. Lett. B265, 239 (1991).

\bibitem{walker}
T. P. Walker et al., Astrophysical J. 376, 51 (1991).

\bibitem{early}
R. A. Harris and L. Stodolsky, Phys. Lett. 116B, 464 (1982); 
Phys. Lett. B78, 313 (1978);
A. Dolgov, Sov. J. Nucl. Phys. 33, 700 (1981);
L. Stodolsky, Phys. Rev. D36, 2273 (1987);
D. Notzold and G. Raffelt, Nucl. Phys. B307, 924 (1988);
P. Langacker, University of Pennsylvania Preprint,
UPR 0401T, September (1989);
R. Barbieri and A. Dolgov, Phys. Lett. B237, 440 (1990);
Nucl. Phys. B349, 743 (1991);
K. Kainulainen, Phys. Lett. B244, 191 (1990);
M. Thomson, Phys. Rev. A45, 2243 (1991);
K. Enqvist, K. Kainulainen and J. Maalampi,
Nucl. Phys. B349, 754 (1991); Phys. Lett. B244, 186 (1991);
K. Enqvist, K. Kainulainen and M. Thomson, Nucl. Phys. 
B 373, 498 (1992);
J. Cline, Phys. Rev. Lett. 68, 3137 (1992);
X. Shi, D. N. Schramm and B. D. Fields, Phys. Rev. D48, 2568 (1993);
G. Raffelt, G. Sigl and L. Stodolsky, Phys. Rev. 
Lett. 70, 2363 (1993);
B. H. J. McKellar and M. J. Thomson, 
Phys. Rev. D49, 2710 (1994);
R. Foot and R. R. Volkas, Phys. Rev. Lett. 75, 4350 (1995). 

\bibitem{pdg} 
Particle data group, C. Caso et al., Eur. Phys. J. C3 (1998).

\bibitem{ekt}
See the paper by Enqvist, Kainulainen and Thomson in Ref.\cite{early}.

\bibitem{raf}
R. E. Lopez, S. Dodelson, A. Heckler and M. S. Turner,
Phys. Rev. Lett. 82, 3952 (1999);
S. Hannestad and G. Raffelt, Phys.  Rev. D 59, 043001 (1999).

\bibitem{rec}
K. Enqvist, K. Kainulainen and A. Sorri, hep-ph/9906452.

\bibitem{shi}
X. Shi, Phys. Rev. D54, 2753 (1996).

\bibitem{pd}
P. DiBari, Private communication.

\end{references}
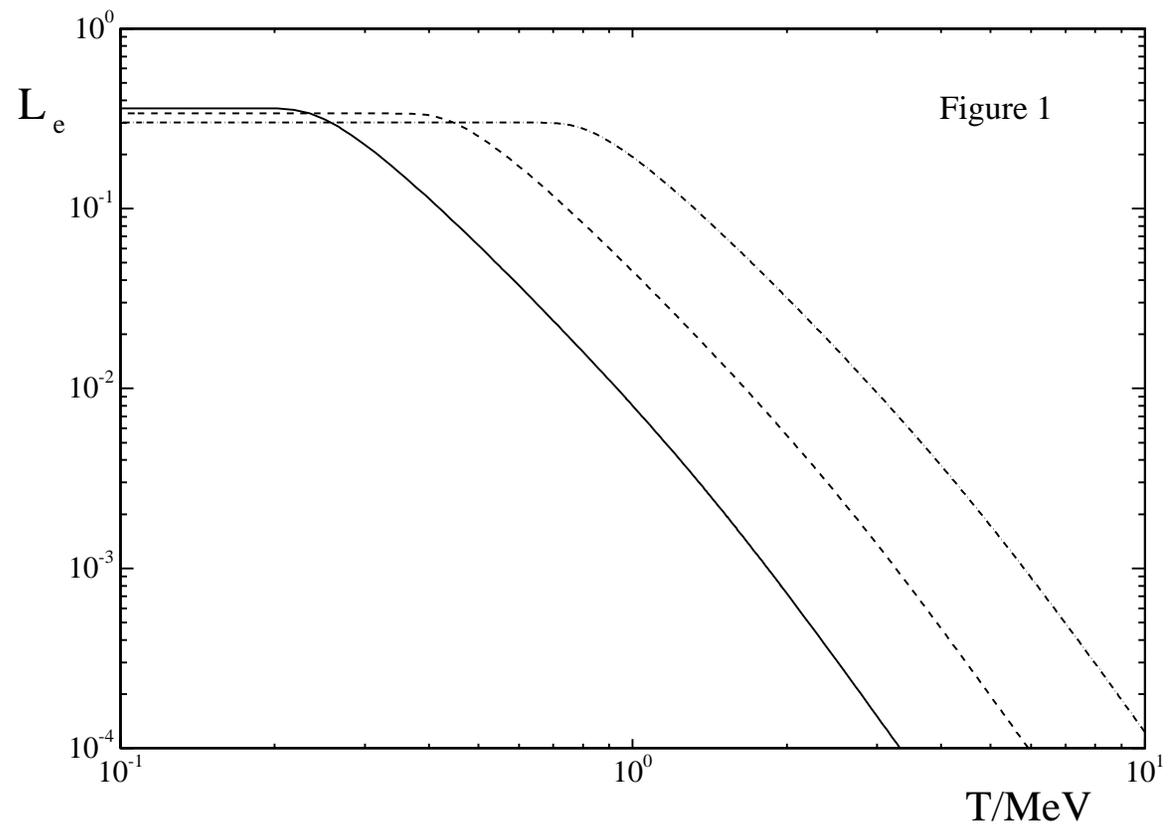
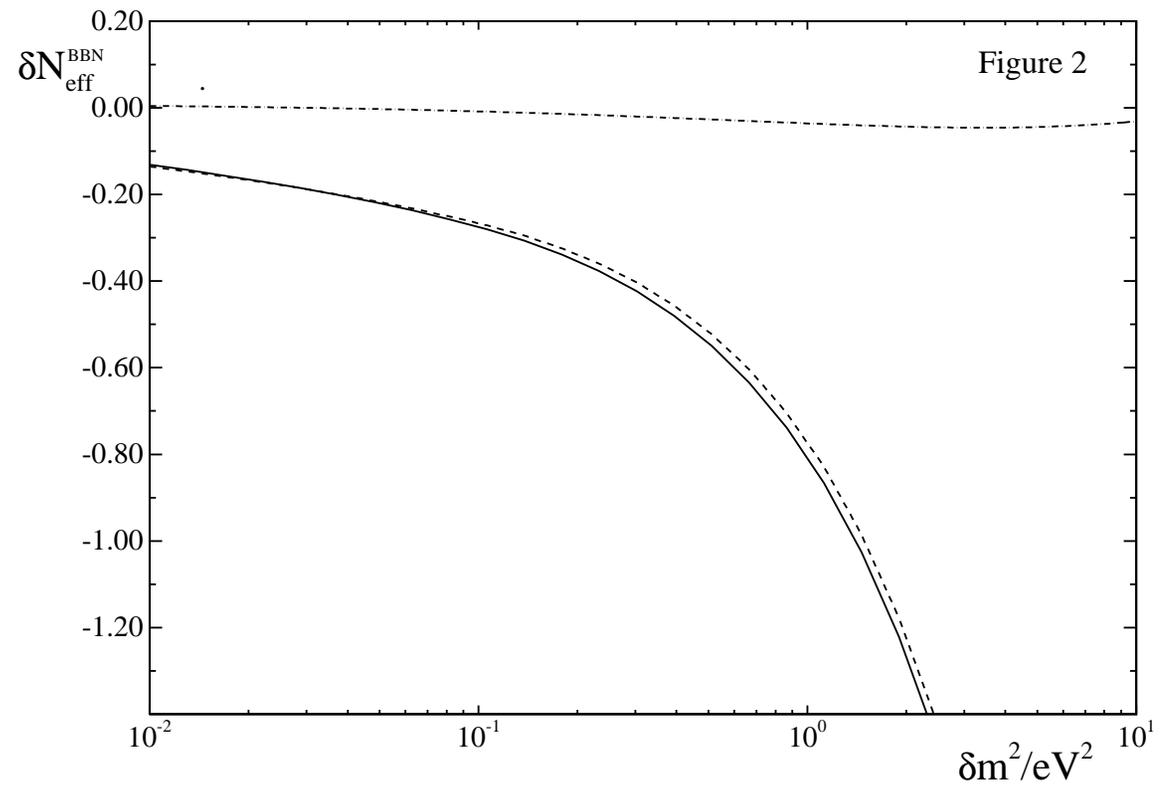
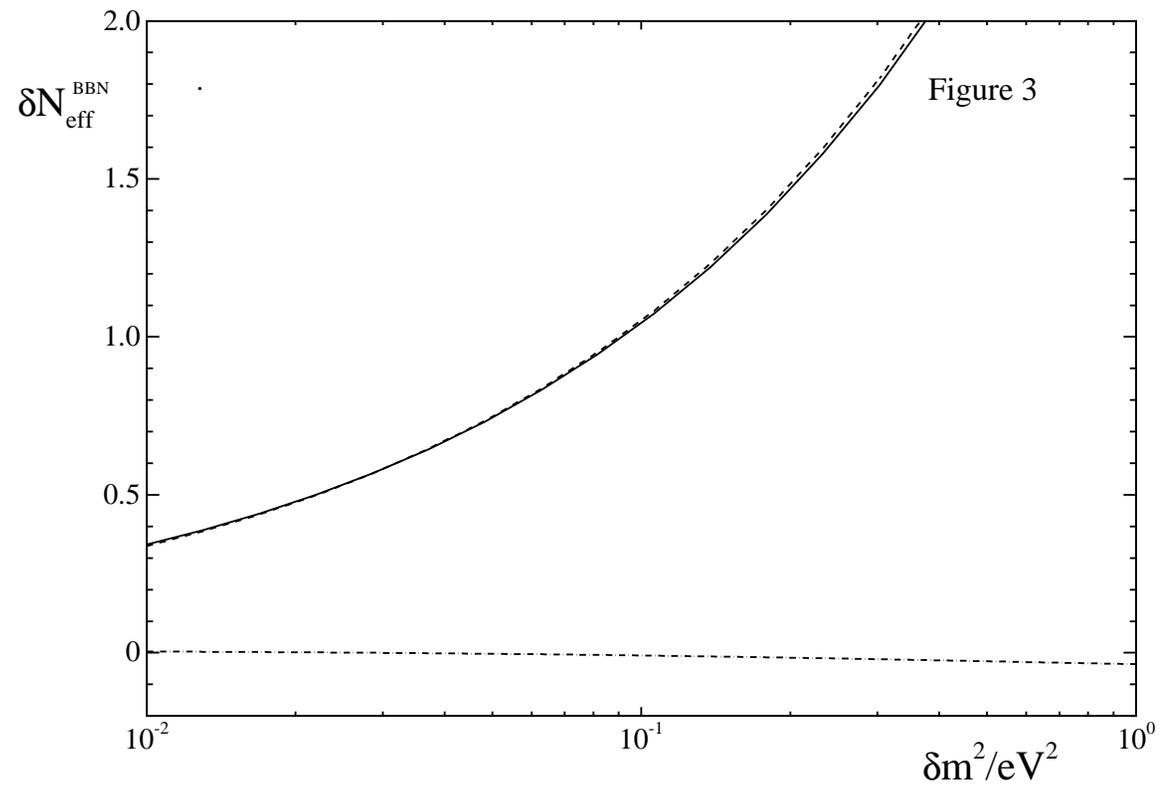
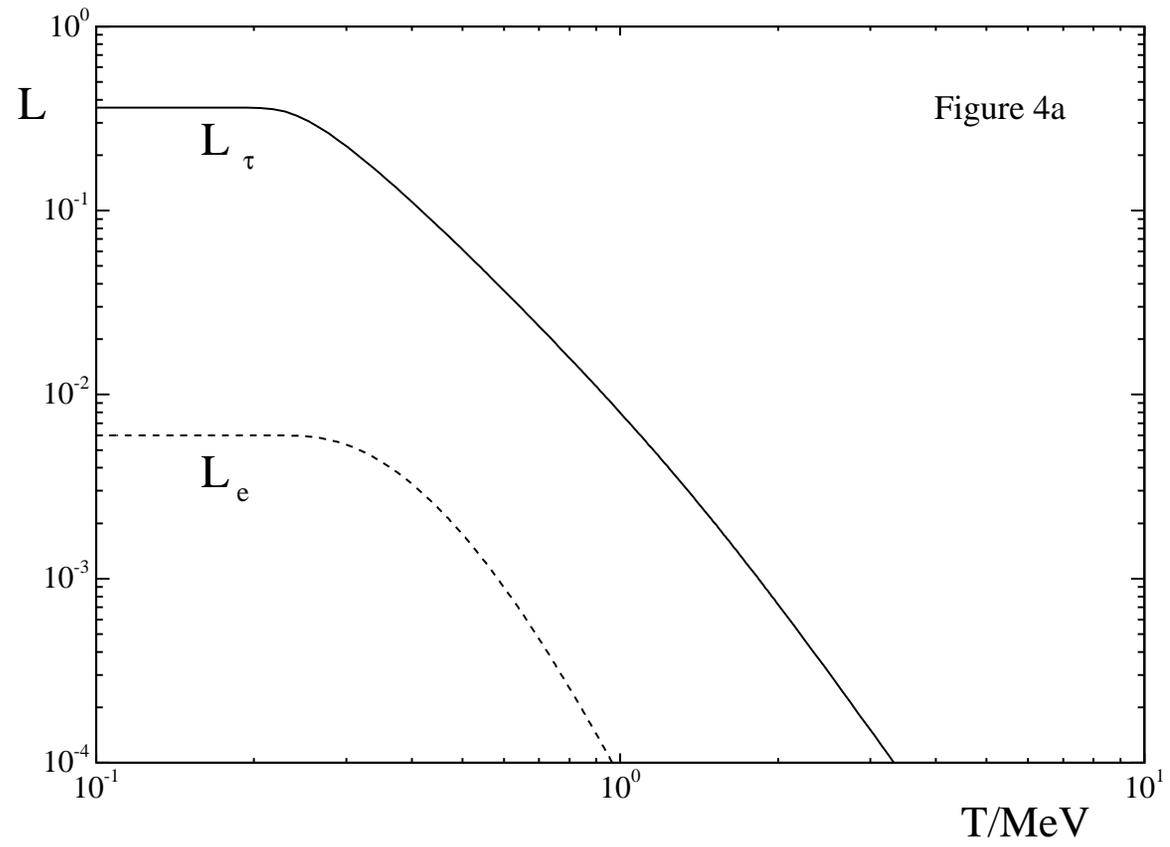
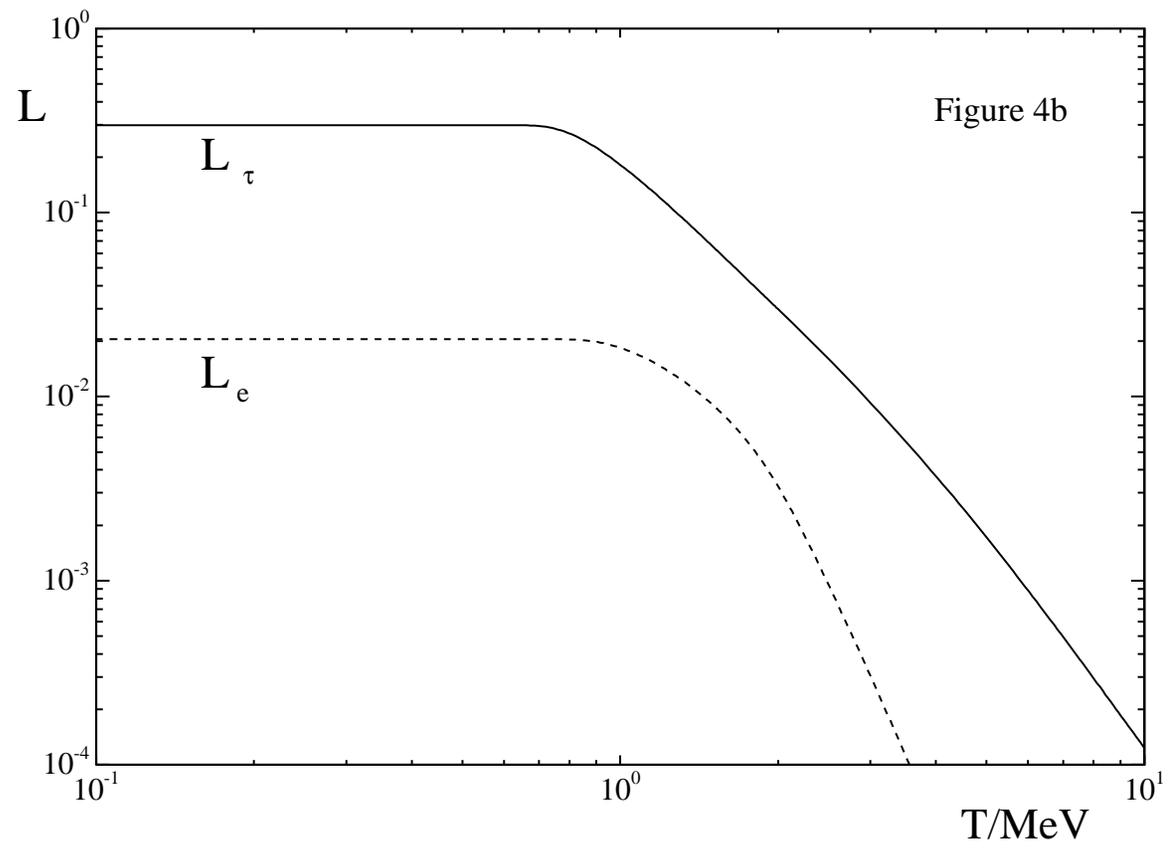
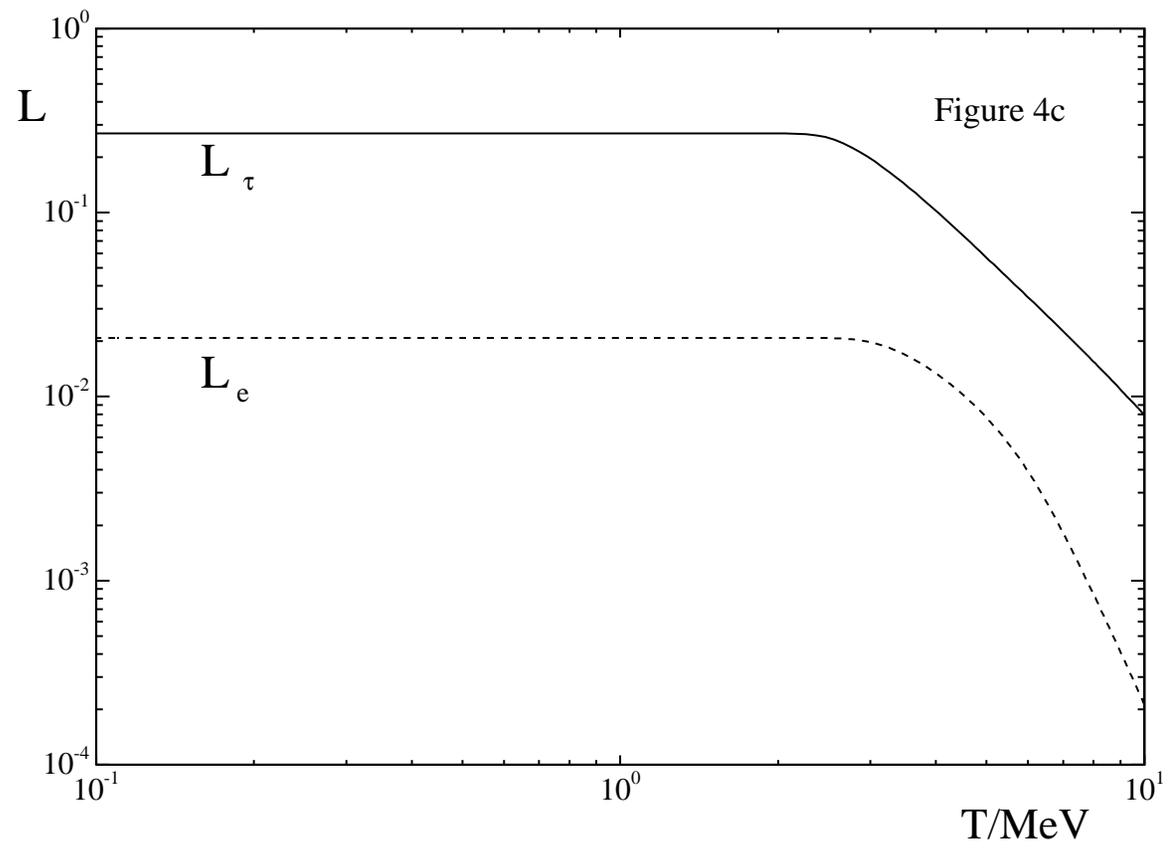
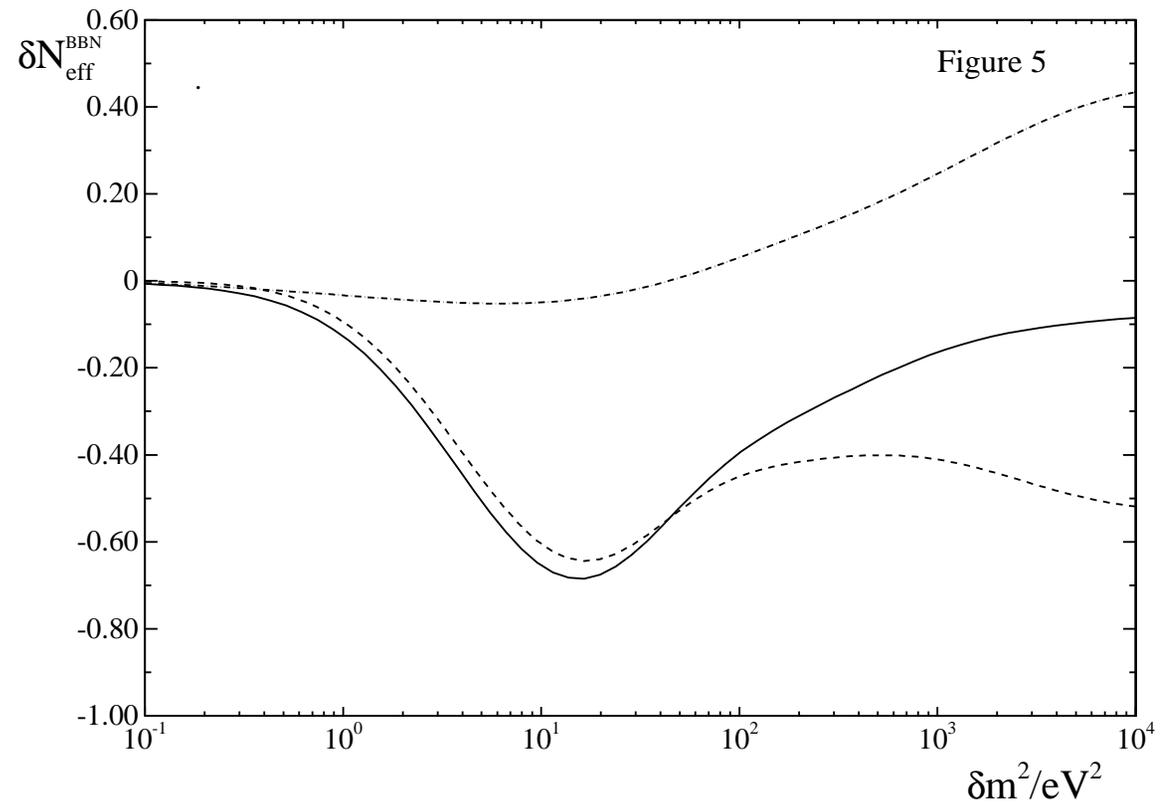
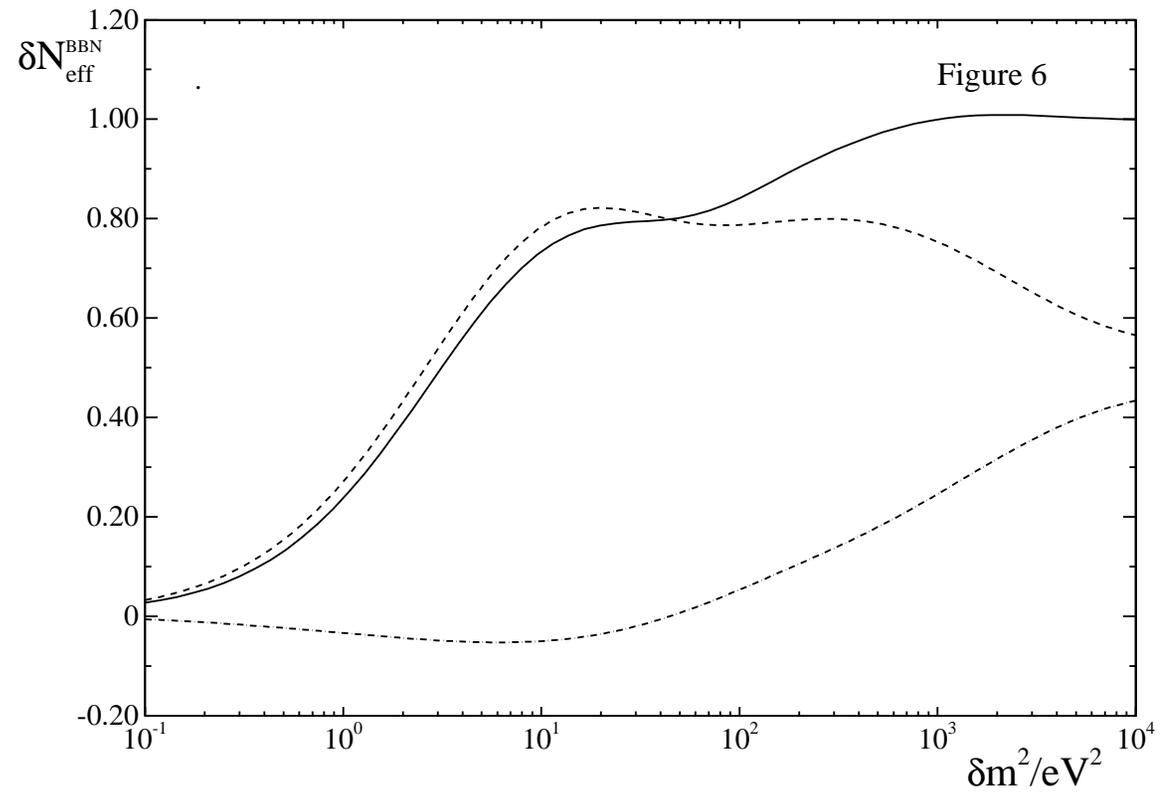
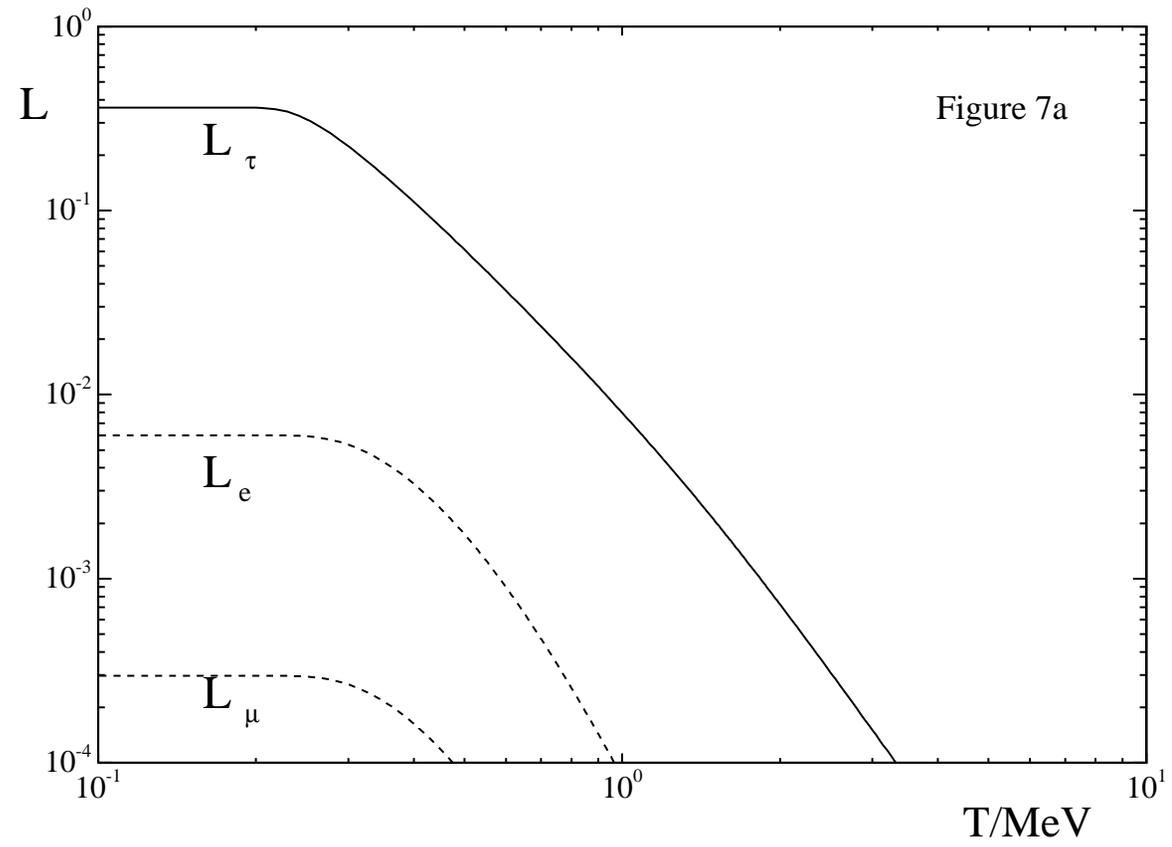
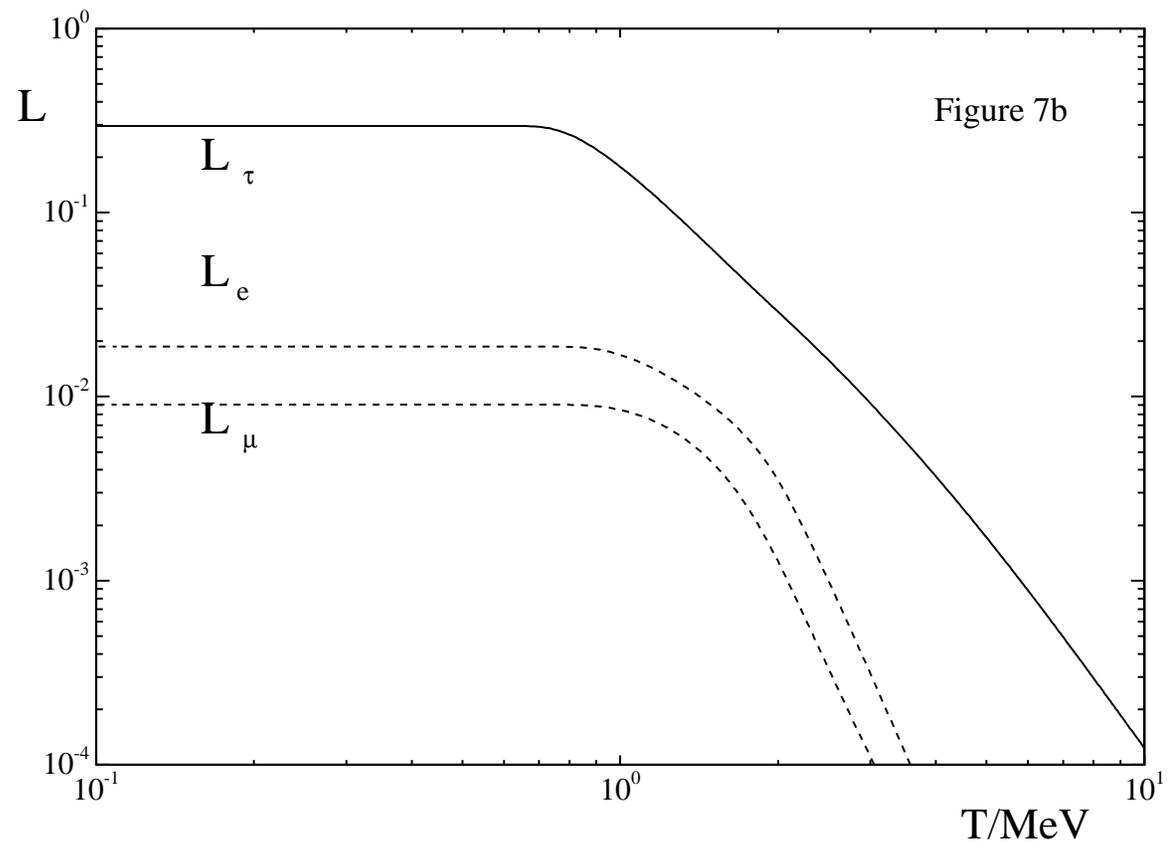
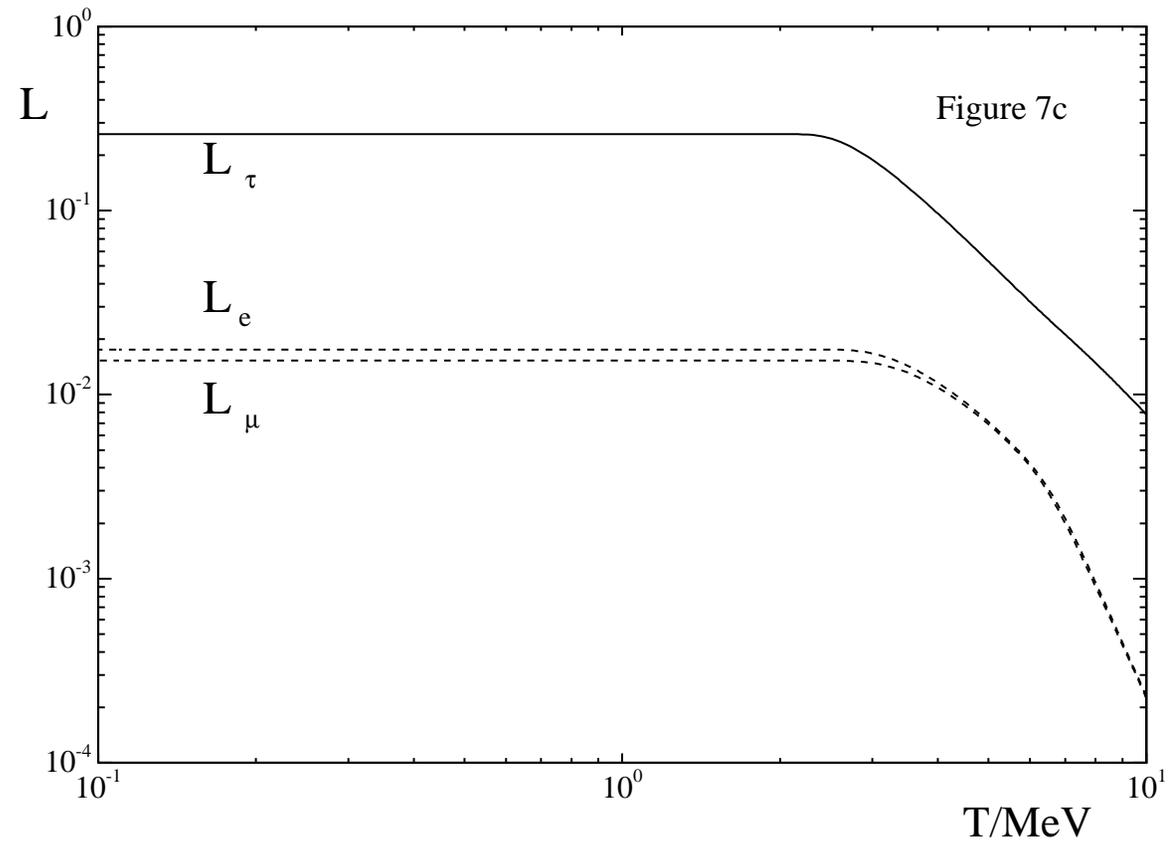
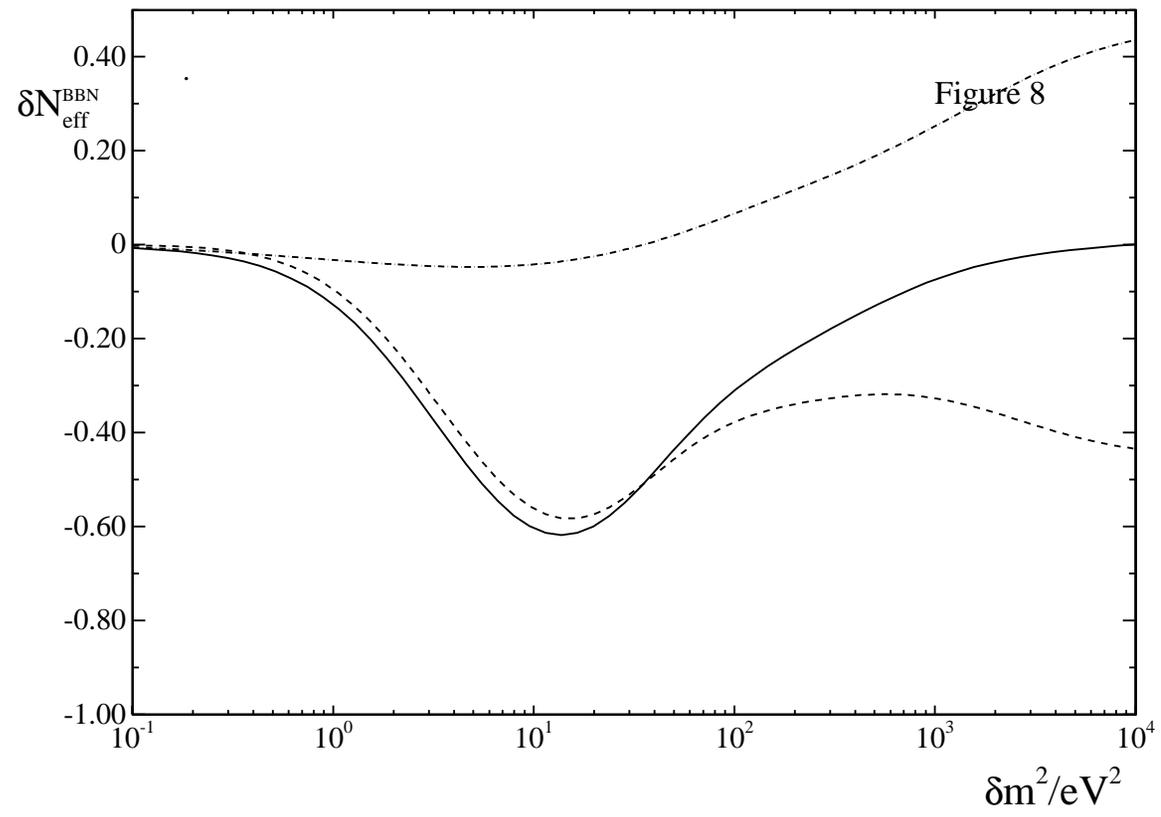
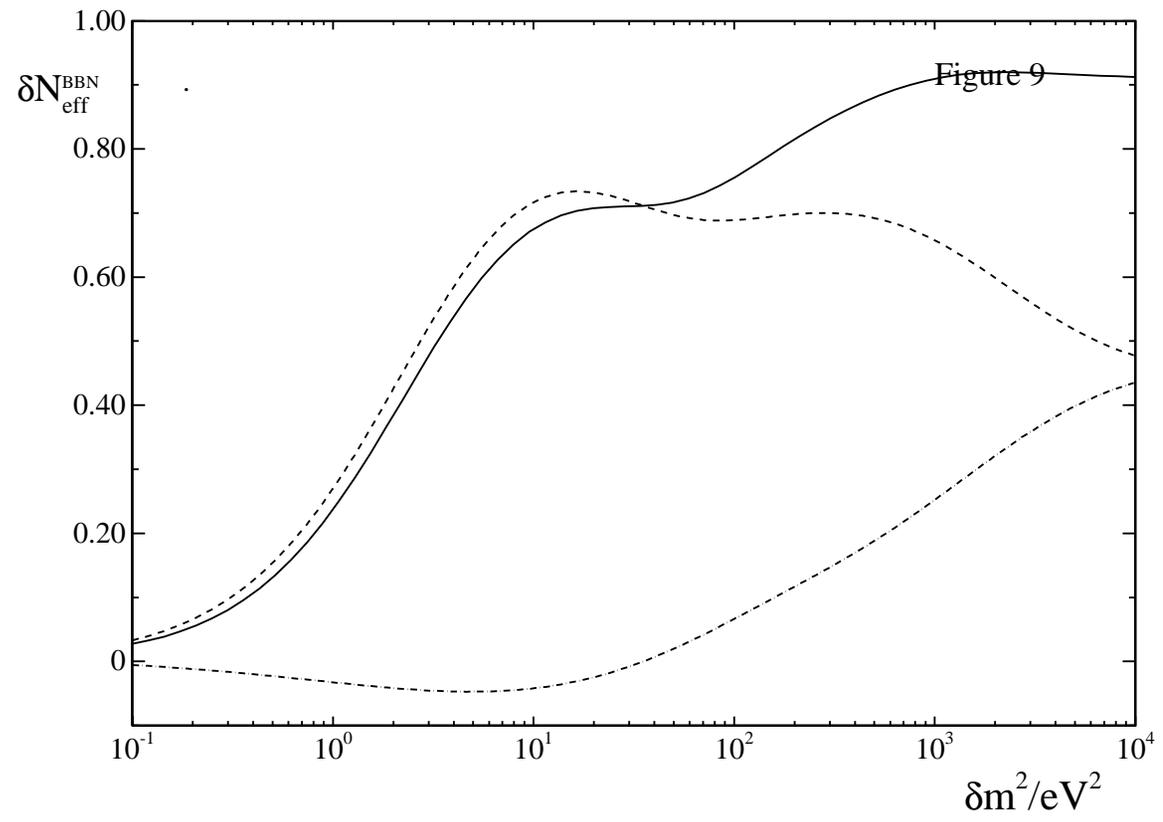
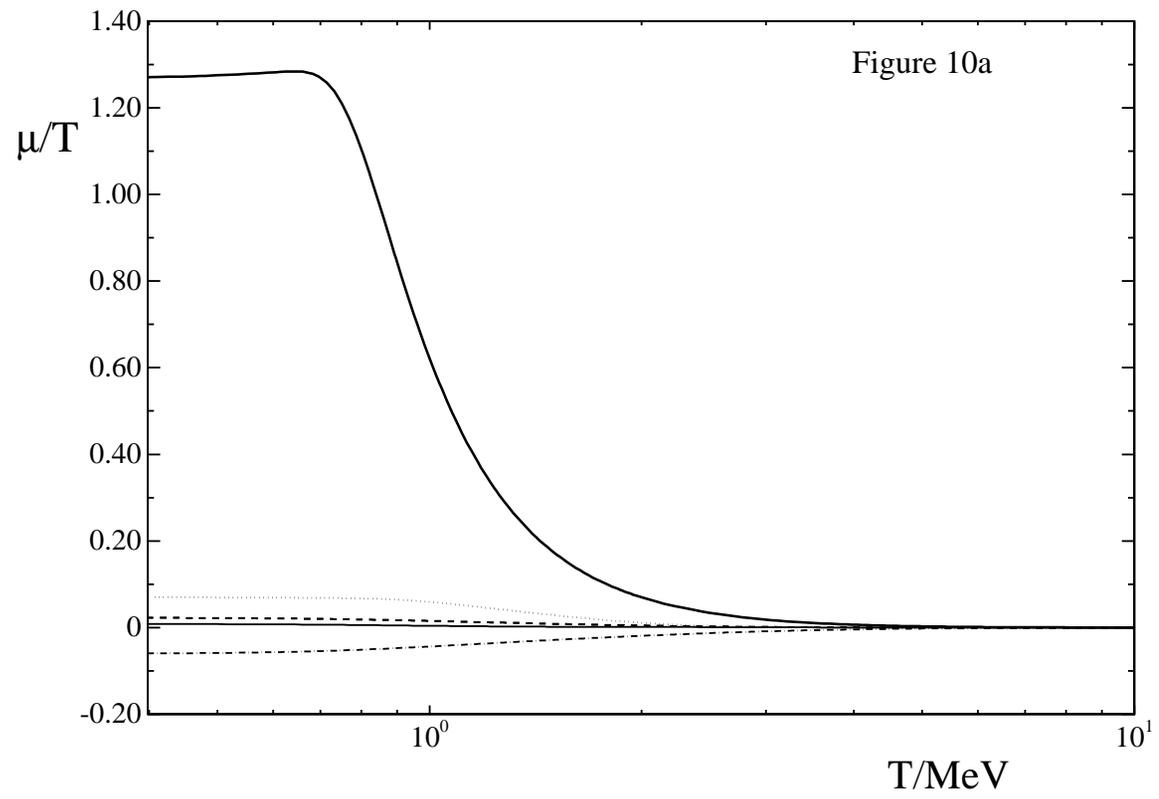
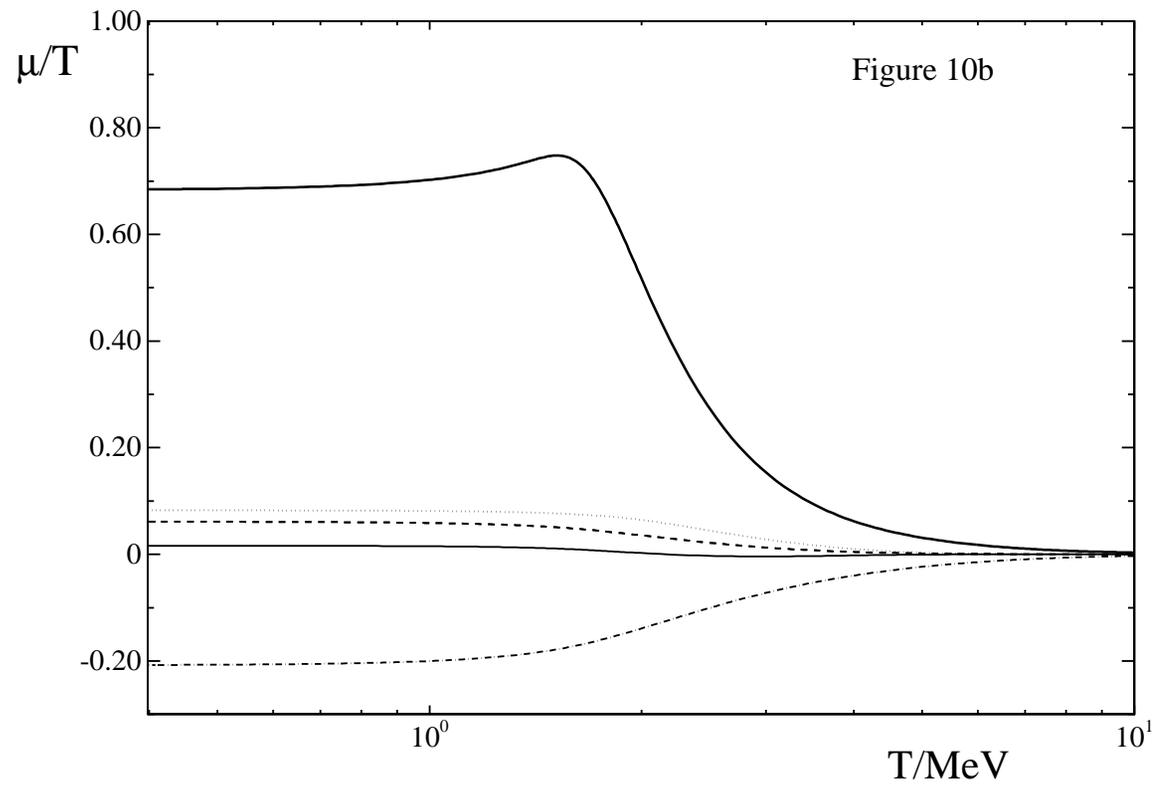
\end{document}